\documentclass[pre,reprint,onecolumn,notitlepage]{revtex4-1}
\usepackage{epsfig,amsmath,amssymb,graphics,color,calc,upgreek}
\usepackage{epsfig}
\usepackage{graphicx}

\newcommand{\Bo}{{\rm Bo}}
\newcommand{\bu}{{\bf u}}
\newcommand{\bx}{{\bf x}}
\newcommand{\bnh}{{\bf \hat{n}}}
\newcommand{\bzh}{\hat{\bf e}_{z}}
\newcommand{\bxh}{\hat{\bf e}_{x}}
\newcommand{\byh}{\hat{\bf e}_{y}}

\newcommand{\pd}[2]{\frac{\partial #1}{\partial #2}}
\newcommand{\der}[2]{\frac{d #1}{d #2}}
\newcommand{\pdd}[2]{\frac{\partial^2 #1}{\partial #2^2}}

\newcommand{\oo}[1]{{\cal O}\!\left( #1 \right)}
\newcommand{\Lc}{L_{\text{circ}}}
\newcommand{\del}{\boldsymbol{\nabla}_s}

\newcommand{\stress}{\boldsymbol{\sigma}}

\newcommand{\ii}{\rm i}
\newcommand{\bcdot}{\boldsymbol{\cdot}}
\newcommand{\bcdotdot}{\boldsymbol{:}}
\newcommand{\bOmega}{\boldsymbol{\Omega}}

\begin{document}

\title{Kinematic irreversibility in surfactant-laden interfaces}

\author{Harishankar Manikantan and Todd M. Squires}

\affiliation{Department of Chemical Engineering\\
University of California, Santa Barbara\\
Santa Barbara, CA 93101, USA}

\begin{abstract}
The surface shear viscosity of an insoluble surfactant monolayer often depends strongly on its surface pressure. Here, we show that a particle moving within a bounded monolayer breaks the kinematic reversibility of low-Reynolds-number flows. The Lorentz reciprocal theorem allows such irreversibilities to be computed without solving the full nonlinear equations, giving the leading-order contribution of surface-pressure-dependent surface viscosity. In particular, we show that a disk translating or rotating near an interfacial boundary experiences a force in the direction perpendicular to that boundary. In unbounded monolayers, coupled modes of motion can also lead to non-intuitive trajectories, which we illustrate using an interfacial analog of the Magnus effect. This perturbative approach can be extended to more complex geometries, and to 2D suspensions more generally. 
\end{abstract}

\maketitle


\section{Introduction}

Understanding the structure, dynamics and rheology of complex interfaces has become increasingly relevant in various fields of biology, technology and industry. High-interface materials are just about everywhere: in suspensions of bubbles, drops, capsules and membranes; in bio-interfaces like the tear film and lung alveoli; and in the industrial processing of coatings, thin films, and foams, to name a few. 

Surface-active materials (`surfactants') play a crucial role in the formation, flow, and stability of these materials. Surfactants lower the surface tension of an interface, and stabilize films against rupture or drops against coalescence. Due to these properties, surfactants are prevalent in the form of soaps, emulsifiers, foaming agents and wetting agents.

Gradients in concentration of such surface species (and therefore, in surface tension) lead to the well-known Marangoni stresses, which give rise to a wide variety of flow phenomena \cite{ScrivenSternling,Leal}. Additional stresses, beyond Marangoni stresses, may be required to deform the surfactant layer. That is, surfactants may exhibit a nontrivial surface rheological response \cite{Fuller2012, Langevin2014}. The impact of interfacial rheology is more subtle than its three-dimensional analog, particularly due to difficulties in clearly demarcating it from other sources of dissipation such as adsorption/desorption \cite{Levich}. Interpretation of measurements is further complicated by surfactant-induced incompressibility \cite{Fischer2004}, which fundamentally alters the boundary conditions obeyed by subphase flow as compared to a clean interface. Additional difficulties are associated with the design of microrheology tools that excite individual modes of deformation. Despite these challenges, recent years have seen the development of techniques \cite{Choi2011, Zell2014, Brooks1999, Verwijlen2011} which have produced clear, quantitative and repeatable results that place interfacial rheology on a firm footing.

For surface rheology to play a nontrivial role, the excess interfacial stress must overwhelm viscous stress from the surrounding bulk phase. This is represented by the Boussinesq number,
\begin{equation}
\Bo=\frac{\eta_s}{\eta L},
\end{equation}
where $\eta_s$ and $\eta$ are surface and bulk shear viscosities of the surfactant and the subphase. $L$ is the length scale over which gradients in surface stresses occur. The classic work of Saffman \& Delbr{\"u}ck \cite{Saffman1975} regarding diffusivity of proteins within lipid membranes, for example, was in the regime of dominant surface viscosity ($\Bo \gg 1$). In this limit, the interfacial layer is governed by the 2D Stokes equations. However, there is no solution to steady flow past a cylinder in two dimensions that satisfies the boundary conditions far from the particle -- the so-called Stokes paradox \cite{Leal}. Saffman \cite{Saffman1976} showed that drag from the subphase ultimately regularizes the divergence at the root of this paradox. This transition from two to three dimensions beyond a Saffman-Delbr{\"u}ck length $\eta_s/\eta$ has since been systematically measured \cite{Prasad2006}. In the current work, we will assume interfacially dominant flows ($\Bo \gg 1$) that completely decouple from the subphase, restricting our attention to distances within the Saffman-Delbr{\"u}ck length and, therefore, to 2D dynamics.

The standard framework for interfacial momentum balance is the Boussinesq-Scriven equation \cite{Scriven1960}, which prescribes a general boundary condition for the bulk flow in the subphase. Viscous stresses within the interface are described by two viscosities, shear ($\eta_s$) and dilatational ($\kappa_s$), much like in bulk fluids, along with Marangoni stresses associated with gradients in surface tension, $\gamma$. Analogous to bulk pressure, one can define a surface pressure, $\Pi$, such that 
\begin{equation}
\Pi(\Gamma)=\gamma_0-\gamma(\Gamma)
\end{equation}
is the surface stress exerted by the surfactant against the surface tension $\gamma_0$ inherent in the clean interface. Here, $\Gamma$ is the local surfactant concentration.

The viscosities of bulk fluids typically change only under extreme pressure, if at all \cite{Bair1998,Hron2001,Rajagopal2003}. Departures from Newtonian solutions have been reported in the contexts of polymer melt processing \cite{Denn1981,Penwell1971} and lubricating oils \cite{Bair1998}, both under high pressures. The surface viscosity of surfactant-laden interfaces, by contrast, depends strongly on the surface concentration (and therefore on $\Pi$) of that surfactant. Surfactant monolayers are much easier to compress than 3D fluids, and even a modest compression can force it undergo phase transitions \cite{Kaganer1999,Fuller2012}. Consequently, surface viscosities change by orders of magnitude over relatively small changes in $\Pi$: in fact, $\eta_s$ often increases exponentially with $\Pi$ \cite{Kurtz2006,Kim2011,Kim2013,Hermans2014}.

In a previous work \cite{Manikantan2017}, we illustrated the non-trivial consequences of surface-pressure-dependent rheology by focusing on lubrication geometries on the interface. These geometries mimic thin gaps between particles (or between physical barriers) embedded within the monolayer. Not only do thin gaps naturally amplify surface pressures and therefore the effects of $\Pi$-dependent viscosity, but the nonlinear governing equations turn out to have a separable solution within this approximation. Among other qualitatively new phenomena, this study revealed that the fluid stresses generated in response to the motion of a probe particle near a confining barrier breaks kinematic reversibility of Stokes flow. Such irreversible dynamics could have far-reaching consequences at the single particle level, such as with interfacial microrheology, or more generally in the macroscopic behavior of 2D suspensions.

In this work, we mathematically quantify the irreversible dynamics that break Newtonian symmetries and potentially affect the design and interpretation of high-$\Bo$ flows. Additionally, we move beyond the thin-gap regime of lubrication theory, and quantify the asymmetric forces and irreversible motion due to $\Pi$-dependent viscosity for arbitrary distances from confining walls. We appeal to asymptotic methods and develop a theory for `small' departures from Newtonian behavior, which we show captures the appropriate limits from the lubrication analysis.

\section{Problem formulation}
\subsection{Governing equations}

\begin{figure}
\centering \includegraphics[width=\textwidth]{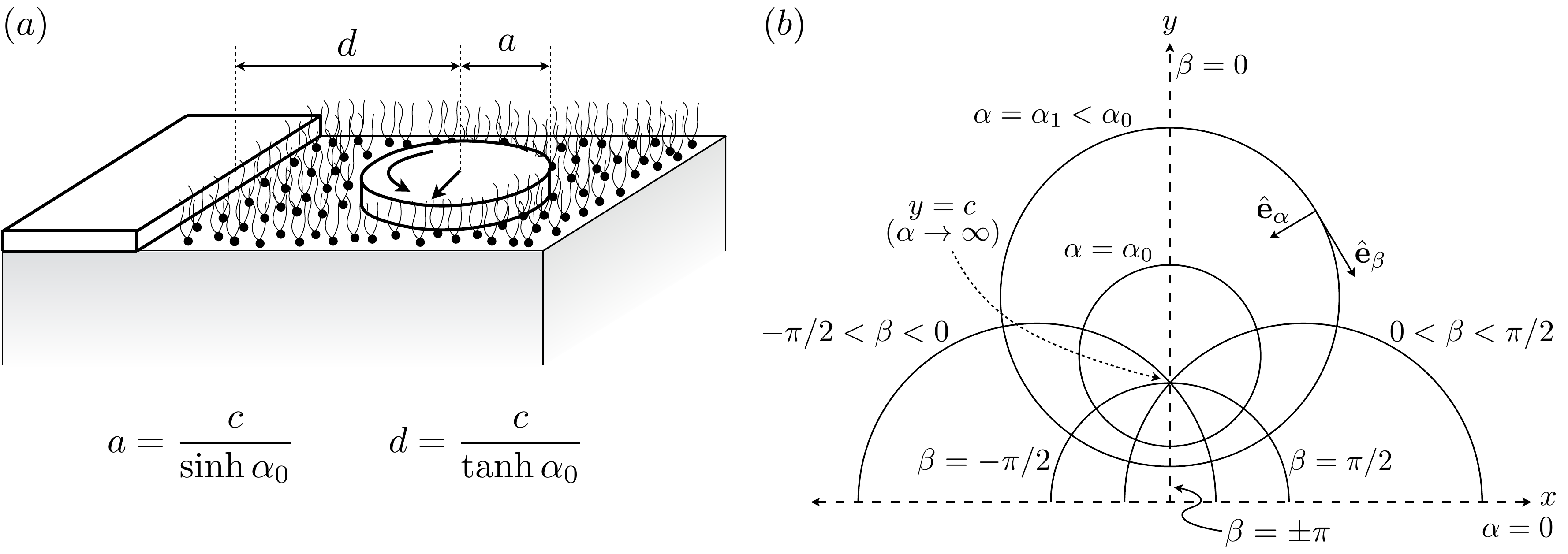}
\caption{$(a)$ Typical geometry of a cylindrical disk embedded on a flat interface near an `interfacial wall'. $(b)$ The bipolar coordinate system. \label{fig:geometry}}
\end{figure}

Let the $x$-$y$ plane represent a surfactant monolayer atop a bulk fluid (Fig.~\ref{fig:geometry}$(a)$). The quasi-steady Boussinesq-Scriven equation describes the interfacial stress balance \cite{Scriven1960}:
\begin{equation} \label{eq:BSgeneral}
\del \Pi = \del \bcdot \left[\eta_s (\del \bu_s +\del \bu_s^T) \right]+  \del \left[(\kappa_s-\eta_s) \del \bcdot \bu_s \right] -  \left.\eta \, \pd{\bu}{z}\right|_{z=0},
\end{equation}
where $\eta_s$ and $\kappa_s$ are the two surface viscosities, and $\eta$ is the shear viscosity of the bulk fluid. The last term couples the interface to the subphase via the bulk viscous traction, which is determined by solving the Stokes equation for the bulk fluid flow $\bu(\bx)$ with Eq.~\eqref{eq:BSgeneral} as the boundary condition. 

While surfactant monolayers are easier to compress than 3D fluids, Marangoni elasticity offers a restoring force against inhomogeneous compression. For example, a translating disk compresses the surfactant monolayer in front of it and dilates the monolayer behind it. This sets up surface concentration gradients that drive Marangoni flows from the front to the back of the disk, returning the surface species to a uniform concentration distribution. When Marangoni stresses are stronger than surface pressure gradients generated by fluid motion, the monolayer behaves like an incompressible 2D fluid \cite{Fischer2004,Manikantan2017}. The surface pressure in Eq.~\eqref{eq:BSgeneral} then acts to enforce incompressibility, while the $\Pi$-dependence of viscosity is explicitly retained.

In the interest of highlighting the effect of $\Pi$-dependent viscosity, we will work in the mathematically simpler regime where $\Bo \gg 1$. Nondimensionalizing Eq.~\eqref{eq:BSgeneral} over characteristic surface viscous stresses and a relevant length scale $L$, the last term is clearly $\oo{1/\Bo}$. While eliminating this coupling re-introduces the divergence associated with 2D Stokes flow at infinity, we will assume that all relevant length scales of interest lie within the Boussinesq length \cite{Prasad2006}. Then, the interface is decoupled from the bulk, and its dynamics follow the Stokes equations,
\begin{equation}\label{eq:BSsimplified}
\del \bcdot \left[\eta_s(\Pi) ( \del \bu_s+\del \bu_s^T)\right] = \del \Pi, \qquad \del \bcdot \bu_s = 0,
\end{equation}
albeit with a surface-pressure-dependent surface shear viscosity. 

For most insoluble surfactants, a moderate change in $\Pi$ drastically changes $\eta_s$. For example, a major constituent of pulmonary surfactant and cell membranes is the phospholipid dipalmitoylphosphatidylcholine (DPPC), which forms stable monolayers at air-water interfaces. Above a critical surface pressure ($\sim 8\, \text{mN}/\text{m}$ at room temperature), DPPC undergoes a phase transition to a liquid-condensed phase \cite{Kaganer1999}, above which its surface viscosity grows exponentially with $\Pi$ \cite{Kim2011,Hermans2014}. This can be qualitatively reasoned in terms of free-area theories of viscosity \cite{Kim2013}, and a good approximation for the surface viscosity of such a `$\Pi$-thickening' monolayer is
\begin{equation}\label{eq:etaPi}
\eta_s(\Pi)=\eta_s^0 e^{\Pi/\Pi_c}.
\end{equation}
Here, $\eta_s^0$ is the surface viscosity at zero surface pressure, and $\Pi_c$ is the pressure scale over which this exponential change happens. For a Newtonian interface, $\Pi_c\rightarrow \infty$, while $\Pi_c$ is measured to be $\sim 6$--$8$ mN/m for DPPC. Certain surfactants (\textit{e. g. } eicosanol \cite{Zell2014}) exhibit `$\Pi$-thinning' due to tilt transitions associated with an increase in surface pressure \cite{Kaganer1999}. All results below are easily adapted to $\Pi$-thinning surfactants by inverting the sign of $\Pi_c$.

Consider the translation and/or rotation of a particle embedded within a bounded surfactant monolayer (Fig.~\ref{fig:geometry}$(a)$). The boundary conditions associated with such motion are
\begin{gather}\label{eq:BC}
\bu_s = {\bf U}_p + \boldsymbol{\Omega}_p\times (\bx-\bx_0) \quad \text{when} \quad \bx \in \ell, \\
\bu_s = {\bf 0} \quad  \text{when} \quad \bx  \in \ell_w,
\end{gather}
where $\ell$ defines the boundary of the particle with center of mass $\bx_0$, and $\ell_w$ represents a stationary `interfacial wall'. ${\bf U}_p$ and $\boldsymbol{\Omega}_p$ are the imposed translational and rotational velocities of the particle.

Clearly, Eq.~\eqref{eq:BSsimplified}, in conjunction with Eq.~\eqref{eq:etaPi}, is extremely nonlinear. We approach this perturbatively, for small departures of $\eta_s(\Pi)$ from $\eta_s^0$. Specifically, when $|\Pi| \ll \Pi_c$, we rewrite the momentum equation in Eq.~\eqref{eq:BSsimplified} as
\begin{equation}
\eta_s^0 \del \bcdot \left[ \left\{ 1 + \frac{\Pi}{\Pi_c} + \oo{\left(\frac{\Pi}{\Pi_c}\right)^2} \right\}( \del\bu_s +\del\bu_s^T )\right] = \del \Pi.
\end{equation}
Nondimensionalizing over characteristic scales of $a$ for length, $U$ for fluid velocity, and $\Pi_0=U\eta_s^0/a$ for surface pressure, and retaining only terms to leading order in the ratio $\Pi/\Pi_c$ gives
\begin{equation}\label{eq:dimless}
\del^* \bcdot  \left[ \left( 1 + \zeta \Pi^* \right) (\del^* \bu^*_s+\del^* \bu^{*T}_s) \right] = \del^* \Pi^*_s.
\end{equation}
For simplicity of notation, we drop the asterisks in dimensionless variables now. We have introduced the dimensionless group 
\begin{equation}
\zeta=\frac{\Pi_0}{\Pi_c}=\frac{U \eta_s^0 }{a \Pi_c} \ll 1,
\end{equation}
which is a ratio of the characteristic surface pressure set up by particle motion to $\Pi_c$. Alternatively, $\zeta$ may be interpreted as a ratio of $U$ to $U_c \sim \Pi_c a /\eta_s^0$, where $U_c$ is the surface velocity characteristic of surface pressure variations comparable to $\Pi_c$. In this sense, $U$ has to be `small' enough for the asymptotic expansion to remain valid. 

Recasting Eq.~\eqref{eq:dimless} in terms of a stress tensor $\stress$, the dimensionless governing equations read:
\begin{subequations}\label{eq:goveqn}
\begin{gather}
\del \bcdot \bu_s = 0, \qquad \del \bcdot \stress = {\bf 0},\\
\stress=-\Pi\, {\bf I} + \del\bu_s +\del\bu_s^T + \zeta\,\Pi(\del\bu_s+\del\bu_s^T ) + \oo{\zeta^2}.
\end{gather}
\end{subequations}
Then, expanding the fluid fields in powers of $\zeta$ as
\begin{equation}
\{ \bu_s, \Pi, \stress \} =\{  \bu_s^{(0)} , \Pi^{(0)}, \stress^{(0)} \}+ \zeta \{ \bu_s^{(1)} , \Pi^{(1)},\stress^{(1)} \}+ \oo{\zeta^2},
\end{equation}
and substituting in Eq.~\eqref{eq:goveqn}, we find the zeroth (Newtonian) order to be 2D Stokes flow, as expected:
\begin{subequations}
\begin{gather}
\del \bcdot \bu_s^{(0)} = 0, \qquad \del \bcdot \stress ^{(0)}= {\bf 0},\\
\stress^{(0)}=-\Pi^{(0)}\, {\bf I} + \del\bu_s^{(0)} +(\del\bu_s^{(0)})^T.
\end{gather}
\end{subequations}
At first order in $\zeta$, however, we have
\begin{subequations}
\begin{gather}
\del \bcdot \bu_s^{(1)} = 0, \qquad \del \bcdot \stress ^{(1)}= {\bf 0},\\
\stress^{(1)}=-\Pi^{(1)} {\bf I} + \del \bu_s^{(1)} +(\del \bu_s^{(1)})^T + \Pi^{(0)}(\del \bu_s^{(0)}+(\del \bu_s^{(0)})^T).
\end{gather}
\end{subequations}
Here, $\stress^{(1)}$ is the stress associated with the leading effect of pressure-dependent rheology. We chose to set up the problem such that the particle is constrained to move with imposed velocities ${\bf U}_p$ and $\boldsymbol{\Omega}_p$, which are entirely accounted for by the leading order flow. Any non-trivial effect of $\Pi$-dependent rheology, therefore, would yield a non-zero force ${\bf F}^{(1)}$ or torque ${\bf T}^{(1)}$ at $\oo{\zeta}$. These can be computed as the moments of the stress tensor $\stress^{(1)}$:
\begin{equation}\label{eq:FB}
{\bf F}^{(1)} = \int \bnh \bcdot \stress^{(1)} d\ell, \qquad {\bf T}^{(1)} = \int \bx \times \stress^{(1)}\bcdot\bnh\, d\ell,
\end{equation}
where $\bnh$ is the normal pointing from the particle into the monolayer.

\subsection{A Reciprocal approach}

Solving the $\oo{\zeta}$ equations of motion is at best tedious. However, the Lorentz reciprocal theorem \cite{KimKarrila} allows us to circumvent solving the inhomogeneous equation, using only solutions to homogeneous Newtonian problems. To illustrate this, first consider an auxiliary problem of the same particle moving in a Newtonian 2D fluid (of constant surface viscosity $\eta_s^0$) with translational and rotational velocities ${\bf U}_{\text{aux}}$ and $\bOmega_{\text{aux}}$, respectively. Let the associated fluid velocity, stress, and hydrodynamics moments on the particle in the auxiliary problem be $\bu_{\text{aux}}$, $\stress_{\text{aux}}$, and $\{ {\bf F}_{\text{aux}},{\bf T}_{\text{aux}}\}$, respectively. Then, we have the vector identity
\begin{equation}\label{eq:rec1}
\del \bcdot (\stress_{\text{aux}} \bcdot \bu_s^{(1)}) = (\del \bcdot \stress_{\text{aux}}) \bcdot \bu_s^{(1)} + \stress_{\text{aux}} \bcdotdot \del \bu_s^{(1)}= \stress_{\text{aux}}  \bcdotdot \del \bu_s^{(1)} ,
\end{equation}
where we have used the momentum equation, $\del \bcdot \stress_{\text{aux}}={\bf 0}$. Similarly,
\begin{equation}\label{eq:rec2}
\del \bcdot (\stress^{(1)} \bcdot \bu_{\text{aux}}) = \stress^{(1)}  \bcdotdot \del \bu_{\text{aux}}.
\end{equation}
Subtracting Eq.~\eqref{eq:rec2} from Eq.~\eqref{eq:rec1} and integrating over the entire fluid domain, we find
\begin{equation}\label{eq:rec}
\int \left[\del \bcdot (\stress_{\text{aux}} \bcdot \bu_s^{(1)}) - \del \bcdot (\stress^{(1)} \bcdot \bu_{\text{aux}}) \right]\,dS = \int \left[ \stress_{\text{aux}}  \bcdotdot (\del \bu_s^{(1)}) - \stress^{(1)}  \bcdotdot \del \bu_{\text{aux}} \right]\,dS.
\end{equation}

The left-hand side of Eq.~\eqref{eq:rec} can be simplified by applying the divergence theorem, and then using the boundary conditions on the disk, from Eq.~\eqref{eq:BC}. Then, applying the force and torque balances from Eq.~\eqref{eq:FB}, the left-hand side becomes
\begin{equation}
\begin{split}
- \int \bnh \bcdot \stress_{\text{aux}} \bcdot \bu_s^{(1)} \,d\ell +& \int  \bnh \bcdot \stress^{(1)} \bcdot \bu_{\text{aux}} \,d\ell \\&= -\left( {\bf F}_{\text{aux}} \bcdot {\bf U}_p^{(1)} +{\bf T}_{\text{aux}} \bcdot \bOmega_p^{(1)} \right) +\left( {\bf F}^{(1)} \bcdot  {\bf U}_{\text{aux}} 
  +{\bf T}^{(1)} \bcdot  \bOmega_{\text{aux}} \right).
 \end{split}
\end{equation} 
We choose to constrain the particle to its leading order motion and determine the external force or torque thus generated at $\oo{\zeta}$; that is, ${\bf U}_p^{(1)} =\bOmega_p^{(1)}={\bf 0}$. Simplifying the right-hand side of Eq.~\eqref{eq:rec} using the definitions of the stresses and putting everything together yields
\begin{equation}\label{eq:correction}
{\bf F}^{(1)} \bcdot  {\bf U}_{\text{aux}}  + {\bf T}^{(1)} \bcdot  \bOmega_{\text{aux}} =- \int \Pi^{(0)}\left[\del\bu_s^{(0)}+(\del\bu_s^{(0)})^T\right]  \bcdotdot \del \bu_{\text{aux}} \,dS.
\end{equation}
Remarkably, ${\bf F}^{(1)}$ or ${\bf T}^{(1)}$ can be evaluated using only the leading-order Newtonian solution $\{\bu^{(0)},\Pi^{(0)}\}$, and the flow field $\bu_{\text{aux}}$ corresponding to another known Newtonian problem in the same geometry. Such a reciprocal formulation is therefore particularly appealing, and we use specific choices of $ {\bf U}_{\text{aux}}$ and $ \bOmega_{\text{aux}}$ in the following sections to determine the $\oo{\zeta}$ forces and torques in model problems. Note that the net force (or torque) at this order scales with the characteristic velocity as $\zeta {\bf F}^{(1)}\sim U^2$: reversing the direction of velocity has no effect on the $\oo{\zeta}$ force (or torque). This departure from Newtonian reversibility and its consequences are the central themes of this work.

\subsection{Newtonian solution in cylindrical bipolar coordinates}

The flow and pressure fields required to evaluate the reciprocal integral in Eq.~\eqref{eq:correction} are all Newtonian solutions to 2D Stokes flow with constant surface viscosity $\eta_s^0$. For cylindrical disks embedded in the monolayer, it is convenient to work in the cylindrical bipolar coordinate system \cite{Wakiya1975,Jeffrey1981,Keh1985} as shown in Fig.~\ref{fig:geometry}$(b)$. In terms of Cartesian coordinates, the bipolar coordinates $(\alpha,\beta)$ are
\begin{equation}
\alpha + \ii \beta = \log \frac{x+\ii (y+c)}{x+\ii (y-c)}.
\end{equation}
Equivalently, $x=h \sin \beta$ and $y=h \sinh \alpha$, where 
\begin{equation}
h=\frac{c}{\cosh \alpha - \cos \beta}
\end{equation}
is a scale factor. Here, $0<\alpha<\infty$, $-\pi\leq \beta < \pi$, and $c$ is a positive number with dimensions of length. Particularly, curves of constant $\alpha$ correspond to non-intersecting, coaxial circles with centers along the $y$-axis. Curves of constant $\beta$ correspond to intersecting circles with centers along the $x$-axis. We set $\alpha=\alpha_0$ to represent a circle of radius $a=c/\sinh \alpha_0$ at a distance $d=c/\tanh\alpha_0$ from the origin along the vertical axis (see Fig.~\ref{fig:geometry}). Conversely, such a circle is described by $\alpha_0=\cosh^{-1}(d/a)$, and then $c^2=d^2-a^2$. Also, we will use $\alpha_1<\alpha_0$ to represent a circle of a larger radius that will serve as the outer wall, such that the monolayer is confined to within $\alpha \in [\alpha_1,\alpha_0]$. The limit of $\alpha_1\rightarrow 0$ corresponds to an outer wall of infinite radius, which represents a plane wall along the $x$-axis.

Defining a stream function $\Psi$ in terms of the fluid velocities such that 
\begin{equation}
u_{\alpha}=-\frac{1}{h}\pd{\Psi}{\beta}, \quad u_{\beta}=\frac{1}{h} \pd{\Psi}{\alpha},
\end{equation}
the Stokes equations can be recast as a biharmonic equation: $\nabla^4\Psi=0$. One way of writing the general solution of this equation that decays sufficiently rapidly at infinity is \cite{Jeffrey1981,Keh1985,Wakiya1975}
\begin{equation}\label{eq:bihar_soln}
\begin{split}
\frac{\Psi}{h}&=A \alpha (\cosh \alpha - \cos \beta) + (B+C\alpha)\sinh\alpha -D\alpha\sin\beta\\
&+ \sum_{n=1}^{\infty}  \left[ \left\{ a_n\cosh(n+1)\alpha + b_n\sinh(n+1)\alpha +c_n\cosh(n-1)\alpha + d_n\sinh(n-1)\alpha   \right\} \cos n\beta \right. \\
&+ \qquad \left. \left\{ a_n'\cosh(n+1)\alpha + b_n'\sinh(n+1)\alpha +c_n'\cosh(n-1)\alpha + d_n'\sinh(n-1)\alpha   \right\} \sin n\beta \right],
\end{split}
\end{equation}
where the coefficients are determined by the general boundary conditions:
\begin{subequations}\label{eq:generalBC}
\begin{align}
\bu^{(0)} &= {\bf U}_p + \boldsymbol{\Omega}_p\times a(-\hat{{\bf e}}_{\alpha} ) \quad \text{at} \quad \alpha=\alpha_0, \\
\bu^{(0)} &= {\bf 0} \quad  \text{at} \quad \alpha=\alpha_1.
\end{align}
\end{subequations}
The Cartesian unit vectors can be represented in terms of the bipolar unit vectors as
\begin{subequations}
\begin{align}
\bxh &= \frac{h}{c}\left[ -(\cosh\alpha \cos\beta -1) ~\hat{\bf e}_{\alpha} -\sinh\alpha \sin \beta ~\hat{\bf e}_{\beta} \right], \\
\byh &= \frac{h}{c}\left[ -\sinh\alpha \sin \beta ~\hat{\bf e}_{\alpha} +(\cosh\alpha \cos\beta -1) ~\hat{\bf e}_{\beta} \right],
\end{align}
\end{subequations}
and $\bzh \times \hat{\bf e}_{\alpha}=\hat{\bf e}_{\beta}$. Using these, and imposing the boundary conditions in Eqs.~\eqref{eq:generalBC}, one can solve for the constants in the expression for the stream function in Eq.~\eqref{eq:bihar_soln}. The general expressions are provided in Appendix \ref{sec:appendix}, and we will use particular cases in the following sections. 

With the stream function known, the pressure field can be obtained from the Cauchy-Riemann equations \cite{Jeffrey1981,Darabaner1967}:
\begin{equation}
\pd{\Pi}{\alpha}=\eta_s^0 \pd{\nabla^2 \Psi}{\beta}, \quad \pd{\Pi}{\beta}=-\eta_s^0 \pd{\nabla^2 \Psi}{\alpha}.
\end{equation}
Then, the Newtonian force (or torque) acting on the disk is obtained by integrating the traction $\bnh\bcdot\stress^{(0)}$ (or its moment) over the cylinder surface \cite{Keh1985}:
\begin{subequations}\label{eq:forcetorque}
\begin{align}
{\bf F}^{(0)}&=4\pi\eta_s^0 (C \bxh + D \byh),\\
{\bf T}^{(0)}&=4\pi\eta_s^0 a (A \sinh\alpha_0 + C \cosh\alpha_0) \bzh,
\end{align}
\end{subequations}
where, again, $\alpha_0$ represents the streamline corresponding to the edge of the disk.

A surprising result \cite{Jeffrey1981,Keh1985}, clear from the expressions for the coefficients $A$, $C$ and $D$, is that translation and rotation are not coupled to each other for a cylinder moving next to a plane wall (\textit{i. e.}, $\alpha_1\rightarrow 0$). For instance, a torque-free cylinder sedimenting parallel to a wall will not rotate -- unlike the corresponding case for a sphere. This is particularly convenient, as rotation and translation can be treated as independent $\oo{1}$ problems in the reciprocal integration for plane walls. We will consider these problems first, which will serve to highlight the nature of the $\oo{\zeta}$ force and its implications in breaking kinematic reversibility. The mathematical machinery is easily extendable to the case of coupled translation and rotation, as we shall then see with an interfacial `journal bearing'. In every case, we first seek the solution to three independent Newtonian problems: a cylinder translating parallel to the wall (${\bf U}_p=U \bxh$), rotating next to the wall ($\boldsymbol{\Omega}_p=\Omega \bzh$), and translating perpendicular to the wall (${\bf V}_p=V \byh$).

\section{Cylinder next to a plane wall}\label{sec:planewall}

\subsection{Translation}

\begin{figure}
\includegraphics[width=\textwidth]{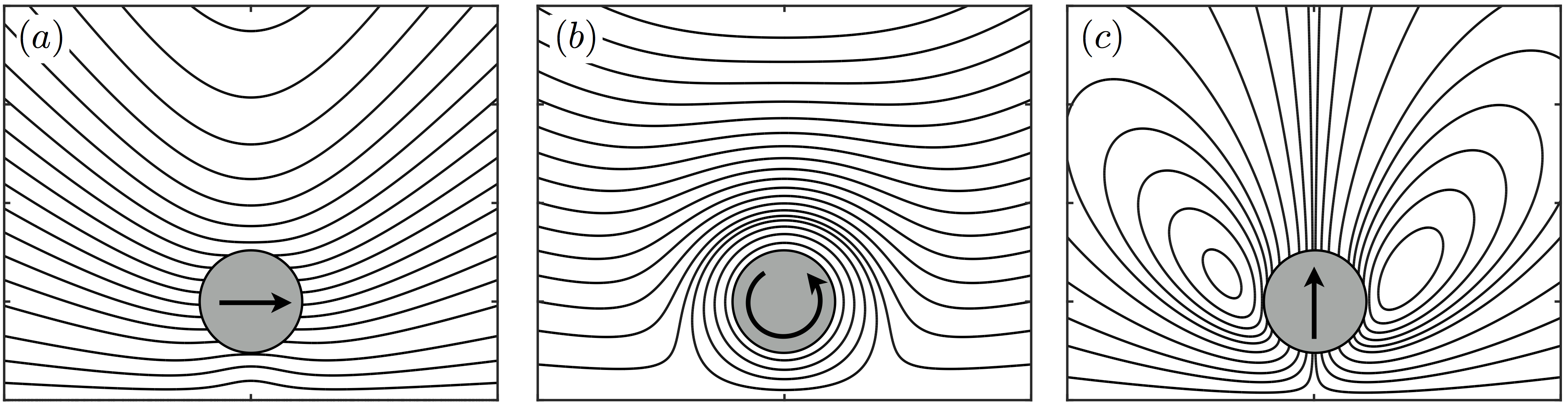}
\caption{The $\oo{1}$ flow fields used to evaluate the lateral forces at $\oo{\zeta}$ for a disk moving next to a plane wall. Streamlines correspond to the Newtonian solutions for a disk $(a)$ translating parallel to wall, $(b)$ rotating adjacent to wall, and $(c)$ translating perpendicular to wall. The plane wall is the bottom edge of each panel. \label{fig:streams}}
\end{figure}

A translating disk compresses the surfactant monolayer in front of it and expands it in the rear. When the interface is incompressible, the surface pressure responds by increasing (or decreasing) in front of (or behind) the disk. Consequently, $\eta_s(\Pi)$ increases and changes the net force on the particle. For a disk moving quasi-steadily along a straight line in an infinite interface, however, the change in $\eta_s$ preserves the dipolar symmetry of the surface pressure distribution, and generates no transverse forces. When near a barrier, however, this symmetry is lost and we expect a finite force in the direction perpendicular to the wall. 

The force and torque on the disk at $\oo{\zeta}$ can be obtained by appropriately choosing the auxiliary Newtonian solution in the reciprocal integration. Specifically, considering separately the cases $\{{\bf U}_{\text{aux}}, \bOmega_{\text{aux}}\}=\{\bxh,{\bf 0}\}$, $\{\byh,{\bf 0}\}$, and $\{{\bf 0},\bzh\}$, Eq.~\eqref{eq:correction} can be rewritten in bipolar coordinates as
\begin{subequations}\label{eq:recip_bipolar1}
\begin{equation}\label{eq:recip_parr1}
F^{\parallel}_{\parallel} ={\bf F}_{\parallel}^{(1)}\bcdot \bxh=- \int_{-\pi}^{\pi} \int_0^{\alpha_0}\! \Pi_{\parallel}\left[\del\bu_{\parallel}+(\del\bu_{\parallel})^T\right]  \bcdotdot \del \bu_{\parallel} \,h^2\, d\alpha \,d\beta,
\end{equation}
\begin{equation}\label{eq:recip_parr2}
F^{\perp}_{\parallel} ={\bf F}_{\parallel}^{(1)}\bcdot \byh=- \int_{-\pi}^{\pi} \int_0^{\alpha_0}\! \Pi_{\parallel}\left[\del\bu_{\parallel}+(\del\bu_{\parallel})^T\right]  \bcdotdot \del \bu_{\perp} \,h^2\, d\alpha \,d\beta,
\end{equation}
\begin{equation}\label{eq:recip_parr3}
T_{\parallel} ={\bf T}_{\parallel}^{(1)}\bcdot \bzh=- \int_{-\pi}^{\pi} \int_0^{\alpha_0}\! \Pi_{\parallel}\left[\del\bu_{\parallel}+(\del\bu_{\parallel})^T\right]  \bcdotdot \del \bu_{\circlearrowright} \,h^2\, d\alpha \,d\beta.
\end{equation}
\end{subequations}
The subscripts $\parallel$, $\perp$, and $\circlearrowright$ denote the velocity and pressure fields corresponding to translation parallel to the wall, perpendicular to the wall, and rotation adjacent to the wall, respectively. The superscripts on the forces denote their directions.

The Newtonian velocity and pressure fields $\{\bu_{\parallel}(\bx),\Pi_{\parallel}(\bx) \}$ are the $\oo{1}$ solution to translation parallel to the wall with velocity $U$. Simplifying the stream-function coefficients from Appendix \ref{sec:appendix}, we find that the only non-zero constants for translation parallel to a plane wall are
\begin{equation}\label{eq:trans_coeff}
A=-B=2b_1=\frac{U}{\alpha_0}\coth\alpha_0,\qquad C=2a_1=-2c_1=-\frac{U}{\alpha_0}.
\end{equation}
The corresponding surface pressure distribution is then
\begin{equation}
\Pi_{\parallel}=\frac{\eta_s^0U}{c} \frac{\coth\alpha_0}{\alpha_0 } \left[ 2\sin \beta \cosh\alpha    +\sin 2\beta(\tanh \alpha_0 \sinh 2\alpha - \cosh 2\alpha)   \right].
\end{equation}
For the auxiliary flow field in Eq.~\eqref{eq:recip_parr1}, the coefficients are the same as above. The only non-zero coefficients for the auxiliary flow field $\bu_{\perp}(\bx)$ in Eq.~\eqref{eq:recip_parr2} (corresponding to a disk moving at velocity $V$ perpendicular to the wall) are
\begin{equation}\label{eq:perp_coeff}
D=2b_1'=-\frac{V}{\alpha_0-\tanh\alpha_0},\qquad a_1'=-c_1'=-\frac{1}{2}D\tanh\alpha_0.
\end{equation}
Similarly, for the auxiliary solution $\bu_{\circlearrowright}(\bx)$ in Eq.~\eqref{eq:recip_parr3} corresponding to rotation at angular velocity $\Omega$, we have
\begin{equation}\label{eq:rot_coeff}
A=-B=2b_1=-a\Omega \frac{\coth \alpha_0}{\sinh \alpha_0},\qquad a_1=-c_1=a\Omega \frac{1}{2\sinh \alpha_0}.
\end{equation}

We nondimensionalize all variables as before, and set $V=U=a\Omega=1$. The integrals in Eq.~\eqref{eq:recip_bipolar1} are cumbersome, but analytically tractable in terms of the streamline $\alpha_0$ and the scale factor $c$ that define a particular cylinder. It is illustrative to work with the dimensionless distance $\delta=d/a$ of the center of the cylinder from the wall, such that $\alpha_0=\cosh^{-1}\delta$ and $c/a=\sqrt{\delta^2-1}$. 

The only non-zero integral at $\oo{\zeta}$ corresponds to Eq.~\eqref{eq:recip_parr2}, and results in a force in the direction perpendicular to the wall:
\begin{equation}\label{eq:fparallel}
F_{\parallel}^{\perp} (\delta)= \frac{2\pi  (2\delta \cosh^{-1}\delta-(1+\delta^2) \sqrt{\delta^2-1} )}{(\delta^2-1)^{3/2} (\cosh^{-1}\delta)^2 (\sqrt{\delta^2-1} -\delta \cosh^{-1}\delta) }.
\end{equation}
Both $F^{\parallel}_{\parallel}$ and $T_{\parallel}$ are identically zero at $\oo{\zeta}$: corrections to the wall-parallel drag or torque arise at best at $\oo{\zeta^2}$. This can be rationalized in terms of the symmetries of the stresses around the inner disk. At $\oo{1}$, fluid stresses are odd-symmetric about the disk along $x$, and add up to a net wall-parallel drag as expected. At $\oo{\zeta}$, however, the stresses are even-symmetric around the disk, and the net result is only a force perpendicular to the wall. To clarify this, consider the net force on the disk:
\begin{equation}
{\bf F}_{\rm net} =  F_d\, \bxh + \zeta (F^{\parallel}_{\parallel} \bxh + F^{\perp}_{\parallel} \byh) + \oo{\zeta^2},
\end{equation}
where $F_d$ is the Newtonian drag from Eq.~\eqref{eq:forcetorque}. In dimensional terms, this reads
\begin{equation}
{\bf F}_{\rm net} =  -\frac{4 \pi \eta_s^0 a U}{\cosh^{-1}(d/a)} \bxh + \frac{(\eta_s^0 U)^2}{\Pi_c a} (F^{\parallel}_{\parallel} \bxh + F^{\perp}_{\parallel} \byh) + \oo{U^3}.
\end{equation}
The $\oo{\zeta}$ corrections scale as $U^2$ and do not change sign upon reversing the direction of motion. Drag against translation (${\bf F}_{\rm net}\bcdot\bxh$) must be an odd function of $U$ to reflect this change of sign, and therefore $F^{\parallel}_{\parallel}=0$. Corrections to wall-parallel drag or torque can thus appear at best at $\oo{U^3}$, or equivalently, at $\oo{\zeta^2}$. This argument is illustrated further in Section \ref{sec:lub}, where the stresses in a thin film are quantified to demonstrate these symmetries.

\begin{figure}
\begin{center}
\includegraphics[width=0.6\textwidth]{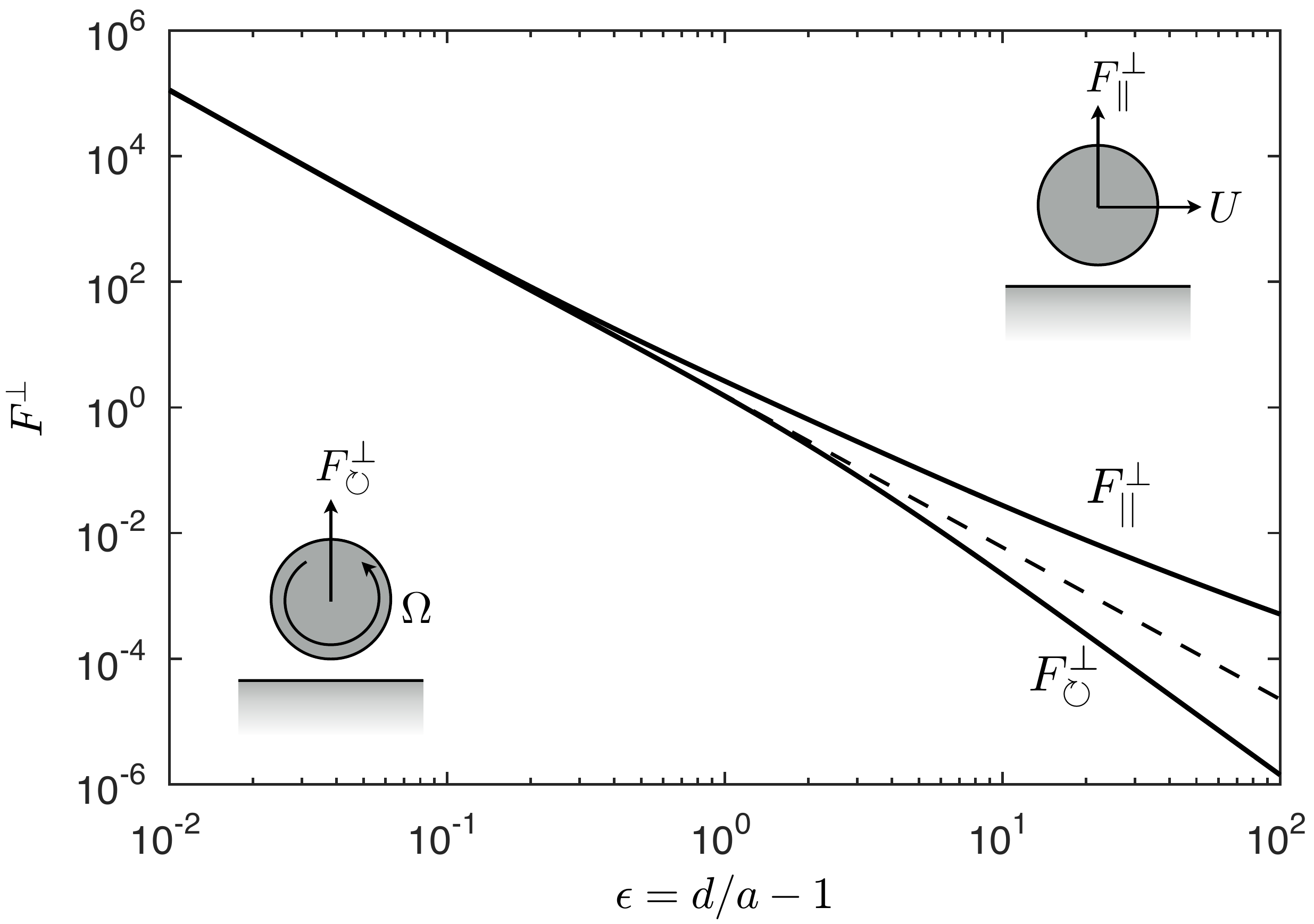}
\end{center}
\caption{Force on the disk due to $\Pi$-dependent $\eta_s$ as a function of distance from the plane wall, following Eqs.~\eqref{eq:fparallel} and \eqref{eq:frotation}. This $\oo{\zeta}$ force acts perpendicular to the wall and breaks the kinematic reversibility of Stokes flow. The dashed line is the $\epsilon^{-5/2}$ behavior predicted by the lubrication analysis (Section \ref{sec:lub}). \label{fig:forcedist}}
\end{figure}

The force $F_{\parallel}^{\perp}$ is shown in Fig.~\ref{fig:forcedist} as a function of the separation $\epsilon=\delta-1$. For large distances from the wall, $F_{\parallel}^{\perp}$ decays to zero. This is expected from our discussion above regarding a cylinder in an infinite quiescent fluid, where symmetry dictates that the $\oo{\zeta}$ contributions be zero. The more useful limit is when the disk is very close to the wall, whereupon Eq.~\eqref{eq:fparallel} behaves as
\begin{equation}\label{eq:fparr_limit}
F_{\parallel}^{\perp} (\epsilon \ll 1)\sim \frac{\pi}{2\sqrt{2}}\,\epsilon^{-5/2} +\oo{\epsilon^{-3/2}}.
\end{equation}
The effect of $\Pi$-dependent rheology is, therefore, more prominent as the fluid film between surfaces layer becomes thinner. Within this thin layer, a $\Pi$-dependent lubrication analysis \cite{Manikantan2017} is possible, which, as we show in Section \ref{sec:lub}, retrieves the same limiting behavior as Eq.~\eqref{eq:fparr_limit}.

The complementary case of a disc approaching (or departing) a wall along the $y$-direction is amenable to a simple symmetry argument. Such a flow is left-right symmetric about the disc, and there will not arise any stresses that result in a net force parallel to the wall. Indeed, it can be shown by integrating the corresponding velocity and pressure fields in place of Eq.~\eqref{eq:recip_bipolar1} that this $\oo{\zeta}$ force $F_{\perp}^{\parallel}={\bf F}_{\perp}^{(1)}\bcdot \bxh$ is indeed zero.

\subsection{Rotation}

As with translation, a disk rotating quasi-steadily in an infinite interface would not feel a net force. However, the surface pressure distribution changes in the proximity of a wall, which in turn changes the local surface viscosity to result in a net force perpendicular to	 the wall. 

Consider a cylindrical disk rotating at angular velocity $\Omega$ near a wall (such that the characteristic fluid velocity scale is now $U=a\Omega$).  The only non-zero coefficients in the stream function corresponding to $\bu_{\circlearrowright}(\bx)$ are the same as in Eq.~\eqref{eq:rot_coeff}, giving a surface pressure distribution:
\begin{equation}
\Pi_{\circlearrowright}=-\frac{\eta_s^0 a \omega}{c} \frac{\coth\alpha_0}{\sinh\alpha_0} \left[ 2\sin \beta (\cosh\alpha-\tanh\alpha_0\sinh\alpha)   +\sin 2\beta(\tanh \alpha_0 \sinh 2\alpha - \cosh 2\alpha)  \right].
\end{equation}
Again, consider the auxiliary Newtonian flow field $\bu_{\perp}(\bx)$ by setting $\bOmega_{\text{aux}}={\bf 0}$ and ${\bf U}_{\text{aux}}=\byh$. The only non-trivial reciprocal integral is
\begin{equation}\label{eq:recip_bipolar2}
F^{\perp}_{\circlearrowright} ={\bf F}_{\circlearrowright}^{(1)}\bcdot \byh=- \int_{-\pi}^{\pi} \int_0^{\alpha_0} \Pi_{\circlearrowright}\left[\del\bu_{\circlearrowright}+(\del\bu_{\circlearrowright})^T\right]  \bcdotdot \del \bu_{\perp} \,h^2\, d\alpha \,d\beta,
\end{equation}
where, like before, the subscripts refer to the modes of motion ($\circlearrowright$ for rotation) and $F^{\perp}_{\circlearrowright}$ is the $\oo{\zeta}$ force on the disk in the direction perpendicular to the wall.
 
Performing this integral, and recasting in terms of the distance $\delta=d/a$ from the center of the disk, the dimensionless vertical force at $\oo{\zeta}$ is now
\begin{equation}\label{eq:frotation}
F_{\circlearrowright}^{\perp} (\delta)= \frac{2\pi  (2\delta \cosh^{-1}\delta-(1+\delta^2) \sqrt{\delta^2-1} )}{(\delta^2-1)^{5/2} (\sqrt{\delta^2-1} -\delta \cosh^{-1}\delta) }.
\end{equation}
As shown in Fig.~\ref{fig:forcedist}, $F_{\circlearrowright}^{\perp}$ decays with $\delta$ at a rate faster than due to a translating disk. More interestingly, the near-wall behavior ($\epsilon = \delta-1 \ll 1$) is identical to that of translation. In fact, expanding Eq.~\eqref{eq:frotation} in this limit, we find 
\begin{equation}\label{eq:frot_limit}
F_{\circlearrowright}^{\perp}(\epsilon \ll 1)\sim \frac{\pi}{2\sqrt{2}}\,\epsilon^{-5/2}
\end{equation}
again, suggesting that the problem is indistinguishable from the case of a disc sliding past the wall when the gap is very small.

A similar integral can be performed to evaluate the $\oo{\zeta}$ force parallel to the wall on a rotating disk. The first order problem maintains the symmetries of the Newtonian problem, and the force $F_{\circlearrowright}^{\parallel}={\bf F}_{\circlearrowright}^{(1)}\bcdot \bxh$ evaluates to zero. Therefore, a cylinder rotating next to a wall in a pressure-thickening medium will experience a force solely in the direction perpendicular to the wall -- if unconstrained, the cylinder will translate vertically away from the wall (see Section \ref{sec:traj}) while maintaining its relative horizontal position.

\subsection{The lubrication limit}\label{sec:lub}

Thin gaps on the interface naturally amplify surface pressure fields, and therefore also the effects of $\Pi$-dependent surface viscosity as evinced by the behavior of $F^{\perp}_{\parallel,\Omega}$ when $\epsilon\rightarrow 0$. In thin gaps, momentum transport is predominantly in one direction, and the governing equations simplify significantly. In fact, it is possible \cite{Manikantan2017} to obtain analytical solutions to flows within 2D lubrication geometries with $\Pi$-dependent $\eta_s$ without resorting to asymptotic methods as we have done here.

\begin{figure}
\centering \includegraphics[width=\textwidth]{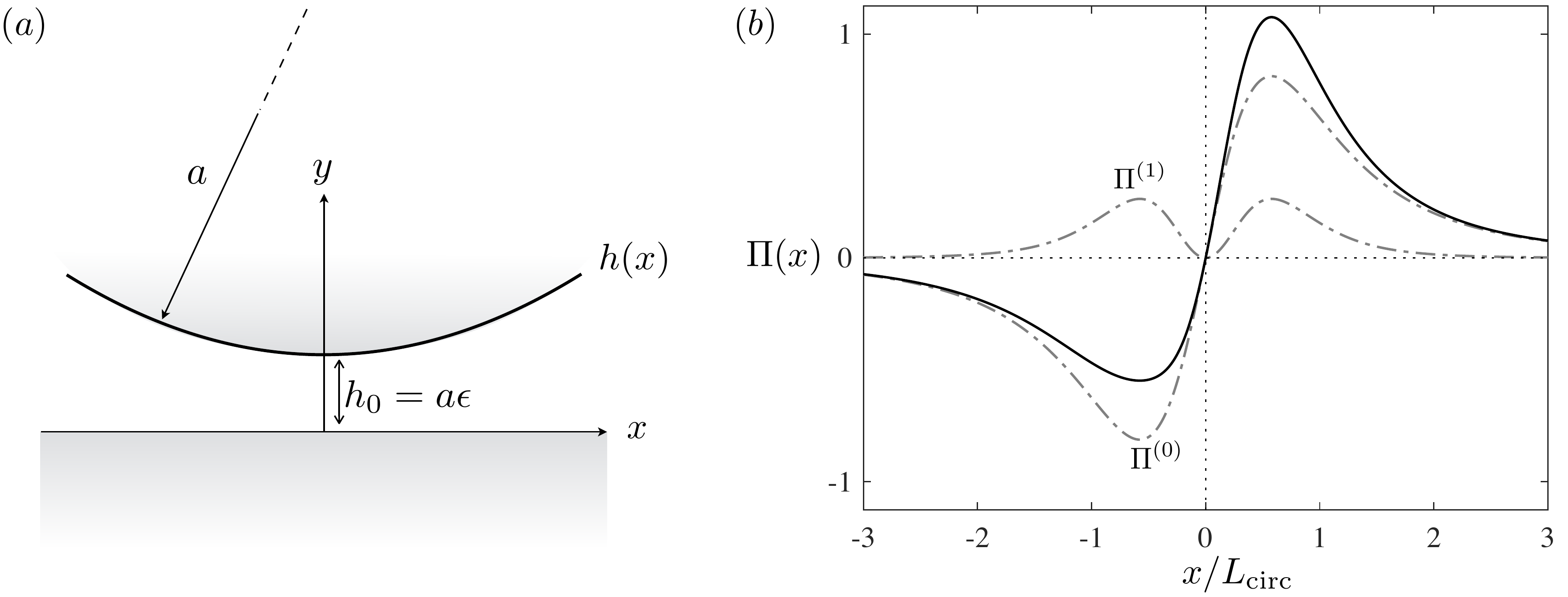}
\caption{$(a)$ Geometry of the thin film between a disk and a wall. $(b)$ Surface pressure distribution (in arbitrary units) in the thin gap (solid line) illustrating the Newtonian and $\oo{\zeta}$ contributions (dash-dot lines) from Eq.~\eqref{eq:Pi_lub_expand}. \label{fig:lub}}
\end{figure}

Let $h(x)$ be the profile of the thin fluid layer between a disk and a wall (see Fig.~\ref{fig:lub}$(a)$) such that the characteristic thickness of this gap, $h_0$, is much smaller that the size of the disk ($h_0 \ll a$). With the same assumptions as before (incompressible 2D flow; high $\Bo$), the momentum equation Eq.~\eqref{eq:BSsimplified} can be shown to simplify in thin gaps to \cite{Manikantan2017}:
\begin{equation}
\pd{\Pi}{x}=\eta_s(\Pi)\pdd{u_s}{y}, \qquad \pd{\Pi}{y}=0.
\end{equation}
These equations take the same form as the Newtonian lubrication theory, albeit with a $\Pi$-dependent viscosity. Surface pressure (and therefore $\eta_s(\Pi)$) is constant along the $y$-direction within this approximation, and therefore the $x$-momentum equation can be integrated like in standard lubrication analysis. In a frame of reference moving with the particle, no-slip conditions are $u_s(x,h(x)) = 0$ on the particle surface and $u_s(x,0) = -U$ at the wall. Then, the fluid velocity field and local flux are
\begin{eqnarray}
u_s(x,y) =  U\left(\frac{ y}{h(x)}-1\right) + \frac{1}{2\eta_s(\Pi)}\der{\Pi}{x} y(y-h(x)), \label{eq:lub_vel} \\
Q  = \int_{0}^{h(x)} u_s(x,y)\,dy = -\frac{U h(x)}{2}  - \frac{h^3(x)}{12\eta_s(\Pi)}\der{\Pi}{x}. \label{eq:lub_flux}
\end{eqnarray}
Despite the strong nonlinearity due to an arbitrary functional form of $\eta_s(\Pi)$, Eq.~\eqref{eq:lub_flux} can be rearranged to a separable ODE:
\begin{equation}\label{eq:lub_separable}
\frac{d\Pi}{\eta_s(\Pi)} = -\frac{12}{h(x)^3}\left(Q + \frac{U h(x)}{2} \right) dx.
\end{equation}
Remarkably, this relation can be integrated (at least in principle) for any given $\eta_s(\Pi)$. Fluid incompressibility dictates that the flow rate $Q$ be constant at every vertical section, and may be obtained by setting the pressure to be equal far away on either side of the particle in Eq.~\eqref{eq:lub_separable}: $\Pi(x \rightarrow -\infty)=\Pi(x \rightarrow +\infty)=0$. We find
\begin{equation}\label{eq:lub_flux2}
Q = -\frac{U}{2} \left. \int_{-\infty}^{\infty} \frac{dx}{h^2(x)} \middle/ \int_{-\infty}^{\infty} \frac{dx}{h^3(x)} \right. .
\end{equation}

Proceeding further needs an explicit expression for the gap profile. When the gap is small compared to the disk radius ($h_0\ll a$), we can approximate the profile as
\begin{equation}
h(x)=h_0(1+x^2/\Lc^2),
\end{equation}
where
\begin{equation}
\Lc=\sqrt{2h_0 a}.
\end{equation}
Then, $Q=-2U h_0/3$, upon which Eq.~\eqref{eq:lub_separable} becomes
\begin{equation}
\int_0^{\Pi(x)}\frac{d\Pi'}{\eta_s(\Pi')}= \frac{6U}{h_0^2} \int_{-\infty}^{x} \left(\frac{4}{3}\frac{1}{(1+x^2/\Lc^2)^3} - \frac{1}{(1+x^2/\Lc^2)^2} \right) dx.
\end{equation}

We now pick the same form of surface viscosity as before: $\eta_s(\Pi)=\eta_s^0 e^{\Pi/\Pi_c}$. Performing the integration, the surface pressure distribution becomes
\begin{equation} \label{eq:Pi_lub}
\Pi(x)=- \Pi_c \ln \left[ 1- \frac{8\eta_s^0 U a^2}{\Pi_c}\frac{x}{(x^2+L_{\text{circ}}^2)^2}  \right].
\end{equation}
The logarithmic dependence imposes an upper bound on $U$: a unique feature of $\Pi$-dependent systems that we discuss at length in a previous work \cite{Manikantan2017}. Here, however, we stay in the regime where the velocity is small enough such that $\zeta \ll 1$. Upon nondimensionalizing Eq.~\eqref{eq:Pi_lub} as before and scaling lengths over $\Lc$, we have
\begin{equation} \label{eq:Pi_lub_nondim}
\Pi(x)=- \frac{1}{\zeta} \ln \left[ 1- \frac{8\zeta a^3}{\Lc^3}\frac{x}{(1+x^2)^2}  \right].
\end{equation}
To compare with our asymptotic analysis from before, we expand Eq.~\eqref{eq:Pi_lub_nondim} in a Taylor series for small $\zeta$:
\begin{equation}\label{eq:Pi_lub_expand}
\Pi(x)= \frac{8 a^3}{\Lc^3}\frac{x}{(1+x^2)^2} + \zeta\,\frac{32  a^6}{\Lc^6}\frac{x^2}{(1+x^2)^4} + \oo{\zeta^2} .
\end{equation}
Within the lubrication approximation, the net vertical force is simply $F^{\perp}=\int \Pi \,dx$. The first term in the expansion in Eq.~\eqref{eq:Pi_lub_expand} corresponds to a Newtonian problem with constant surface viscosity $\eta_s^0$, and the integrated vertical force at this order is zero. This is consistent with kinematic reversibility arguments, and from the shape of the pressure distribution at $\oo{1}$ as shown in Fig.~\ref{fig:lub}$(b)$. The $\oo{\zeta}$ term, however, has a finite contribution resulting in a net `lift' force away from the wall that breaks Stokesian reversibility. Using $\Lc=\sqrt{2h_0 a}$ and $\epsilon=h_0/a$, we find
\begin{equation}\label{eq:lub_force_wall}
F^{\perp}\sim \zeta \, \frac{\pi}{2\sqrt{2}} \epsilon^{-5/2},
\end{equation}
which is exactly the $\epsilon \ll 1$ limit of Eq. \eqref{eq:fparr_limit}. 

This lubrication analysis not only verifies the thin-gap limit of the reciprocal theorem approach, but also illustrates the symmetries of the stresses at $\oo{\zeta}$. The pressure distribution from Eq.~\eqref{eq:Pi_lub_expand} elucidates the origin of the $\oo{\zeta}$ force perpendicular to the wall due to $\Pi$-dependent $\eta_s$, as shown in Fig.~\ref{fig:lub}$(b)$. It is also evident from Fig.~\ref{fig:lub}$(b)$ that the leading-order correction to the drag on the particle in the direction parallel to the wall can occur only at $\oo{\zeta^2}$ due to the even symmetry of $\Pi(x)$ at $\oo{\zeta}$.

\section{Cylinder within a cylinder}\label{sec:JB}

\subsection{The 2D journal bearing}
The framework of the bipolar cylindrical coordinate system and the reciprocal theorem developed here can be easily adapted to more complex geometries. Consider for instance a cylinder rotating within a cylindrical cavity, with a $\Pi$-thickening surfactant occupying the gap between them (Fig.~\ref{fig:spirals}$(a)$). The 2D journal bearing is a well-studied limit of this geometry when the fluid is Newtonian and the characteristic thickness of the fluid gap is small. Although our analysis is not restricted to thin fluid gaps, we shall again use results from lubrication theory \cite{Leal,Manikantan2017} to compare our solutions in the appropriate limits. Notably, translation and rotation are now coupled, unlike in the case of a plane wall.

Mathematically, the problem is identical to the previous sections, with the fluid now confined between $\alpha=\alpha_0$ and $\alpha=\alpha_1$. The inner cylinder, of radius $a$ and identified by the streamline $\alpha_0$, is translated and/or rotated by imposing an external force and/or torque. The external cylinder, of radius $b>a$ corresponding to $\alpha_1 < \alpha_0$, is held stationary.

\begin{figure}
\includegraphics[width=\textwidth]{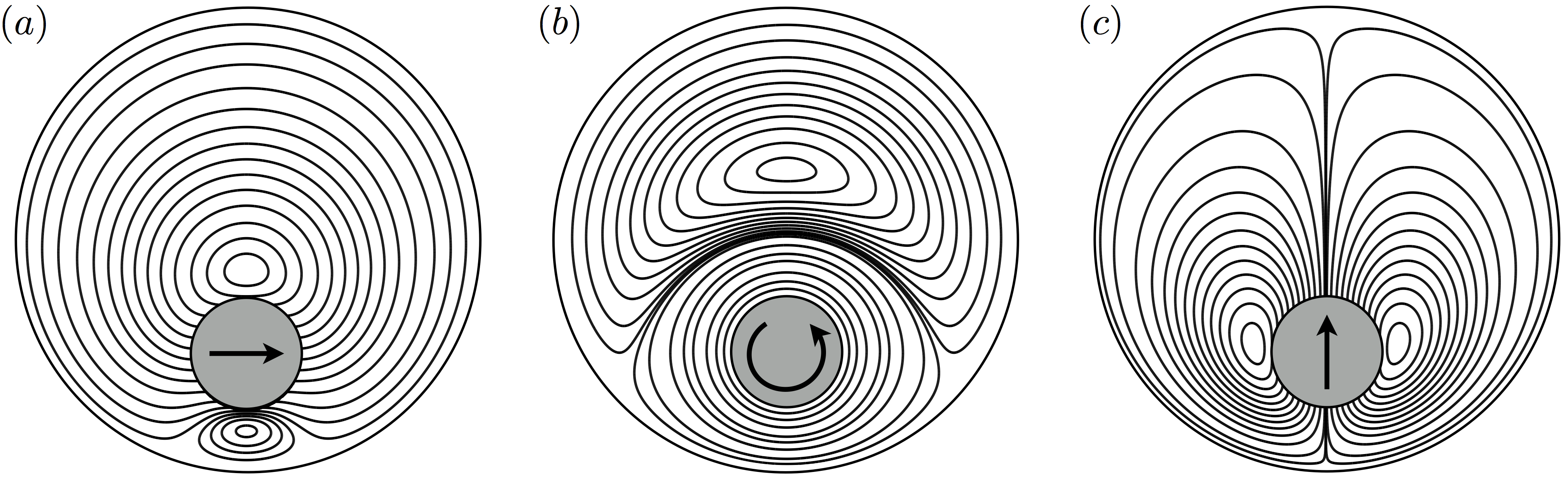}
\caption{Newtonian streamlines corresponding to $(a)$ azimuthal translation of the inner cylinder, $(b)$ rotation, and $(b)$ radial translation. \label{fig:JBstreams}}
\end{figure}

The governing equation at $\oo{1}$ is, again, $\nabla^4\Psi=0$, subject to boundary conditions now applied at a finite $\alpha_1\neq 0$. The coefficients in the stream function in Eq.~\eqref{eq:bihar_soln} follow from the general solution in Appendix \ref{sec:appendix}. The Newtonian streamlines corresponding to different modes of motion of the inner cylinder are shown in Fig.~\ref{fig:JBstreams}.

It is useful to define a clearance $\epsilon$ and eccentricity $\lambda$ such that the radius of the outer cylindrical wall is $b=a(1+\epsilon)$, and the distance between the centers of the cylinders is $\lambda a \epsilon$. Written this way, $\lambda=0$ represents concentric cylinders, whereas the cylinders touch tangentially at $\lambda=1$. These geometric parameters then relate to $\alpha_0$ and $\alpha_1$ as
\begin{equation}
\sinh \alpha_0=\frac{(\lambda +1)(\lambda  \epsilon +\epsilon +2)}{2 \lambda }  \sqrt{\frac{(\lambda -1) (\lambda \epsilon-\epsilon -2)}{(\lambda +1) (\lambda  \epsilon +\epsilon +2)}} ,\qquad \cosh \alpha_1=\frac{\lambda ^2 \epsilon +\epsilon +2}{2 \lambda  (\epsilon +1)}.
\end{equation}
In a Newtonian fluid, kinematic reversibility requires that the radial force (\textit{i. e.}, towards the center of the outer cylinder) on a rotating disk be zero. The net hydrodynamic force generated by the no-slip condition on the inner cylinder, if at all, should act in the azimuthal direction (\textit{i. e.}, parallel to the outer wall). This Newtonian force is particularly significant when the fluid layer between the cylinders is thin, and the methods of lubrication theory may be used to evaluate this `journal-bearing force' \cite{Leal}:
\begin{equation}\label{eq:lub_force_leal}
F_{\text{lub}}=F_{\circlearrowright}^{\parallel}(\epsilon \ll 1)=\frac{12\pi\eta_s^0\Omega a}{\epsilon^2}\left[ \frac{\lambda}{(2+\lambda^2)(1-\lambda^2)^{1/2}} \right].
\end{equation}
The $\epsilon^{-2}$ dependence on the clearance makes it particularly suitable for engineering applications, where a load balances this lubrication force in, say, a shaft rotating within a journal enclosing a thin film of lubricant.

\begin{figure}
\includegraphics[width=\textwidth]{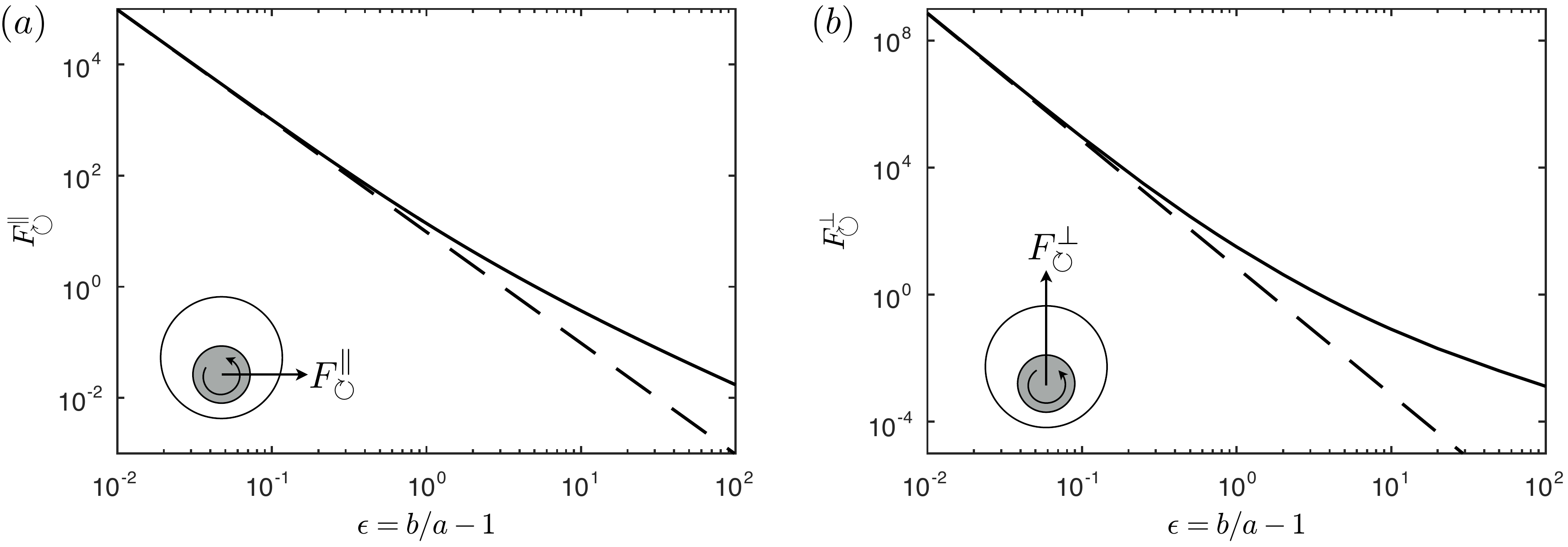}
\caption{Forces on a rotating disk in a journal bearing geometry. $(a)$ The Newtonian journal-bearing force (solid line) at $\oo{1}$ and the prediction from standard lubrication theory (dashed line). $(b)$ The $\oo{\zeta}$ symmetry-breaking force (solid line) from the reciprocal theorem calculation and the prediction from $\Pi$-dependent lubrication theory (dashed line). In both cases, the eccentricity is $\lambda=0.5$.  \label{fig:JBforce}}
\end{figure}

We may readily compare the hydrodynamic force on the inner cylinder from the general Newtonian solution obtained in bipolar coordinates to this lubrication limit in when the gap is thin. Following Eq.~\eqref{eq:forcetorque}, the radial component (or $F_y$) is indeed zero, and the azimuthal component (or $F_x$) is
\begin{equation}\label{eq:lub_force}
F_{\circlearrowright}^{\parallel}= \frac{8\pi\eta_s^0\Omega a \sinh ( \alpha_1) \sinh (\alpha_0- \alpha_1)}{4 \sinh \left(\alpha _0\right) \sinh \left(\alpha _1\right) \sinh \left(\alpha _0-\alpha _1\right)-(\alpha _0-\alpha_1) \left[\cosh \left(2 \alpha _0\right)+\cosh \left(2 \alpha _1\right)-2\right]}.
\end{equation}
This is the Newtonian journal-bearing force without limitations on the thickness of the fluid layer. As expected, this relation limits to zero when $\alpha_1\rightarrow 0$ due to the decoupling of rotation and translation in two-dimensional Stokes flow next to a wall. Rewriting in terms of $\lambda$ and $\epsilon$, we can compare this force against the predictions of lubrication theory (Fig.~\ref{fig:JBforce}$(a)$). When $\epsilon\rightarrow 0$, Eq.~\eqref{eq:lub_force} approaches the prediction from lubrication theory (Eq.~\eqref{eq:lub_force_leal}).

Now consider a fluid with $\Pi$-dependent viscosity filling the gap between the cylinders, as in Fig.~\ref{fig:spirals}$(a)$. Like with the motion of a cylinder near a plane wall, stresses generated in this geometry would result in a net force along the line connecting the centers of the cylinders. We will assume for now that the inner disk is constrained in the horizontal direction, so there is no azimuthal motion. The reciprocal integral then takes the form:
\begin{equation}\label{eq:recip_bipolar3}
F^{\perp}_{\circlearrowright} =- \int_{-\pi}^{\pi} \int_{\alpha_1}^{\alpha_0} \Pi_{\circlearrowright}\left[\del\bu_{\circlearrowright}+(\del\bu_{\circlearrowright})^T\right]  \bcdotdot \del \bu_{\perp} \,h^2\, d\alpha \,d\beta,
\end{equation}
where, like before, the subscripts denote the mode of motion (rotation corresponding to the $\oo{1}$ solution and radial translation as the auxiliary solution), and the superscript on $F$ denotes that this force is perpendicular to the outer wall.

The stream functions corresponding to this geometry are more involved than before, and a closed-form expression for the integral in Eq.~\eqref{eq:recip_bipolar3} eludes us. However, a numerical integration is straightforward and $F_{\circlearrowright}^{\perp}$ is shown in Fig.~\ref{fig:JBforce}$(b)$. As with translation/rotation near a plane wall, this regime can be interpreted to represent `small' velocities such that the corresponding surface pressure is much smaller than $\Pi_c$. On the other hand, a separable solution valid for thin films but for arbitrary values of $\zeta$ can be found in a manner similar to Section \ref{sec:lub}. The correction to the Newtonian pressure distribution again breaks the symmetry, generating a `lift' force that is, to leading order \cite{Manikantan2017},
\begin{equation}\label{eq:lub_force_limit}
F_{\circlearrowright}^{\perp} (\epsilon \ll 1) \sim \frac{18 \pi \zeta}{\epsilon^4} \frac{\lambda^3(3-2\lambda^2 )}{(2+\lambda^2)^2(1-\lambda^2)^{5/2}}.
\end{equation}
As shown in Fig.~\ref{fig:JBforce}$(b)$, these two calculations are consistent in the regime where both the angular velocity (and therefore, $\zeta$) and fluid layer thickness (or $\epsilon$) are small.

\subsection{Irreversible trajectories}\label{sec:traj}
Both $F_{\circlearrowright}^{\parallel}$ and $F_{\circlearrowright}^{\perp}$ above were evaluated for a disk torqued such that it rotates at a constant angular velocity $\Omega$ at a fixed azimuthal location. If we relax this constraint on its position, one can expect the disk to move in both the radial (due to the $O(1)$ Newtonian force $F_{\circlearrowright}^{\parallel}$) and the azimuthal (due to the $\oo{\zeta}$ force $F_{\circlearrowright}^{\perp}$) directions. The net effect is a unique spiral trajectory that brings the inner cylinder closer to the center of the outer cylinder in a $\Pi$-thickening fluid. 

In order to obtain these trajectories, we consider a force-free inner cylinder imposed with an external torque $T^e$. Translation and rotation are now coupled, and we assume that the inner cylinder moves with angular velocity $\Omega$, azimuthal velocity $U$, and radial velocity $V$ towards the center of the outer cylinder, all of which are expanded as
\begin{equation}
\{U,V,\Omega\}=\{U^{(0)},V^{(0)},\Omega^{(0)}\}+ \zeta\{U^{(1)},V^{(1)},\Omega^{(1)}\} +\oo{\zeta^2}.
\end{equation}

At $\oo{1}$, the journal-bearing force $F_{\circlearrowright}^{\parallel}$ balances the hydrodynamic drag to azimuthal motion with velocity $U^{(0)}$. Additionally, rotation as well as wall-parallel translation are both resisted by hydrodynamic torques, the sum of which should balance the applied torque $T^e$. It is also clear from the stream function coefficients that radial motion is decoupled from azimuthal and rotational motion, and therefore $V^{(0)}=0$. The force and torque balances at $O(1)$ are:
\begin{equation}\label{eq:U0}
\begin{pmatrix}
\mathcal{R}_{\parallel}^F & \mathcal{R}_{\circlearrowright}^F \\[0.5em] \mathcal{R}_{\parallel}^T & \mathcal{R}_{\circlearrowright}^T
\end{pmatrix}
\begin{pmatrix}
U^{(0)} \\[0.5em] \Omega^{(0)}
\end{pmatrix}
=
\begin{pmatrix}
0 \\[0.5em] -T_e
\end{pmatrix}.
\end{equation}
Here, $\mathcal{R}$ is the resistance coefficient, with the superscript denoting the hydrodynamic moment (force or torque), and the subscript representing the mode of motion (wall-parallel translation or rotation). The $\oo{1}$ flow fields are Newtonian, and these coefficients can be determined from Eq.~\eqref{eq:forcetorque} and the stream-function coefficients in Appendix \ref{sec:appendix} to give $U^{(0)}$ and $\Omega^{(0)}$.

These $\oo{1}$ modes of motion generate Newtonian fluid fields $\{\bu_{\parallel},\Pi_{\parallel} \}$ or $\{\bu_{\circlearrowright},\Pi_{\circlearrowright} \}$ that in turn could result in $\oo{\zeta}$ forces and torque on the inner cylinder. These corrections ${\bf F}^{(1)}$ and ${\bf T}^{(1)}$ can be determined by appropriately choosing the auxiliary flow field similar to Eq.~\eqref{eq:recip_bipolar1}. Stresses at this order follow the same symmetries along lines of constant $\alpha$ as with the case of a disk near a plane wall, and integrate to a non-zero force only in the direction perpendicular to the outer wall:
\begin{equation}\label{eq:recip_bipolar4}
F^{\perp}_1=- \int_{-\pi}^{\pi} \int_{\alpha_1}^{\alpha_0} \left(\Pi_{\circlearrowright} + \Pi_{\parallel}\right)  \left[\del \left(\bu_{\circlearrowright}+\bu_{\parallel} \right)+\left(\del \left(\bu_{\circlearrowright} +\bu_{\parallel} \right)\right)^T\right]  \bcdotdot \del \bu_{\perp} \,h^2\, d\alpha \,d\beta,
\end{equation}
where $F^{\perp}_1={\bf F}^{(1)}\bcdot\byh$. Similar integrations show that $F^{\parallel}_1={\bf F}^{(1)}\bcdot\bxh$ and $T_1={\bf T}^{(1)}\bcdot\bzh$ are identically zero. As with a plane wall, corrections to $U^{(0)}$ and $\Omega^{(0)}$ arise at best at $\oo{\zeta^2}$.

\begin{figure}
\centering
\includegraphics[width=\textwidth]{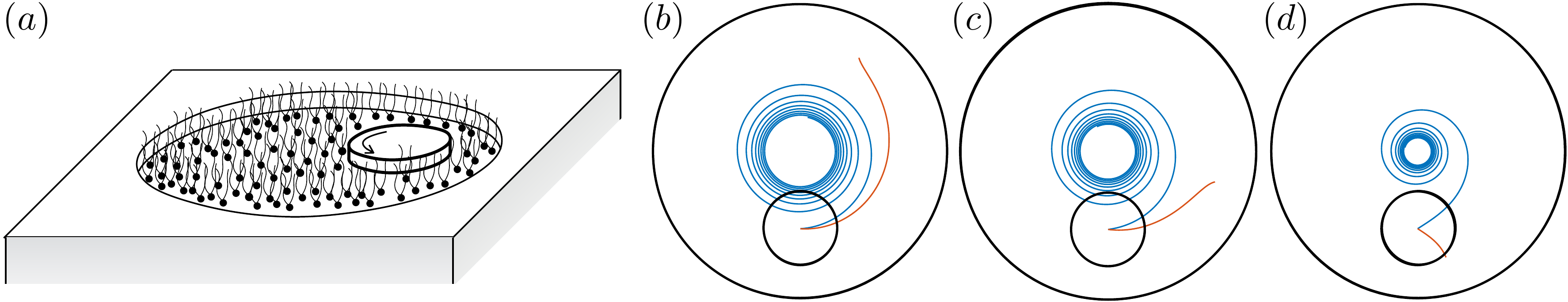}
\caption{$(a)$ The interfacial journal bearing. When the disk is torqued, it spirals inward (or outward) in a $\Pi$-thickening (or -thinning) surfactant. $(b)$--$(d)$ Trajectories obtained by time marching Eq.~\eqref{eq:spirals} to a finite time (when spiraling inward, in blue) or until the disk touches the outer wall (when spiraling outward, in red). The three cases correspond to different values of $\zeta$: $(b)$~$|\zeta|=0.05$, $(c)$~$|\zeta|=0.1$, and $(d)$~$|\zeta|=0.5$. \label{fig:spirals}}
\end{figure}	

If the disk is not constrained, this force acts to push it closer to the center of the outer cylinder. Like before, vertical translation decouples from rotation and horizontal translation. The $\oo{\zeta}$ vertical force balance is simply
\begin{equation}\label{eq:Vcorr}
\mathcal{R}_{\perp}^{F} V^{(1)}=-F^{\perp}_1.
\end{equation}
Between Eqs.~\eqref{eq:U0} and \eqref{eq:Vcorr}, the translational velocities of the inner disk are
\begin{equation}\label{eq:spirals}
U\approx U^{(0)}+ \oo{\zeta^2},\qquad V\approx \zeta V^{(1)} + \oo{\zeta^2}.
\end{equation}
Time-marching using these velocities, we obtain the trajectories in Fig.~\ref{fig:spirals}. As hypothesized, stresses generated at $\oo{\zeta}$ in a $\Pi$-thickening surfactant lead to a spiral trajectory inward. When the surfactant is $\Pi$-thinning ($\zeta\rightarrow-\zeta$), these spiral trajectories bring the disk closer to the outer wall. 

Indeed, when $\alpha_1\rightarrow 0$, rotation and wall-parallel translation decouple, and the calculations are simple enough to track analytically. The result is that a disk rotating adjacent to a plane wall drifts away from the wall (assuming positive $\zeta$ for $\Pi$-thickening) with a velocity $V\approx \zeta V^{(1)}$ with $V^{(1)}$ as in Eq.~\eqref{eq:Vcorr}. The $\oo{\zeta}$ force then is $F^{\perp}_1(\alpha_1=0)=F^{\perp}_{\circlearrowright}$ as given in Eq.~\eqref{eq:frotation}. The following limits can be easily obtained:
\begin{equation}
V(\delta\rightarrow 1) \sim \frac{\zeta}{12 (\delta -1)}, \qquad V(\delta \rightarrow \infty) \sim \frac{\zeta}{2 \delta^3},
\end{equation}
where, like before, $\delta=d/a$ is the dimensionless distance of the center of the disk from the wall and $V$ is in units of the tangential velocity $a\Omega$. 

These transverse velocities are within experimentally accessible regimes for microfabricated ferromagnetic probes \cite{Choi2011,Zell2014} of radius $a\sim 50\,\upmu \text{m}$ on a DPPC monolayer ($\Pi_c\sim 8\,\text{mN}/\text{m}$ and $\eta_s^0\sim10\,\upmu \text{Ns}/\text{m}$). Obviously, thin gaps amplify the effect due to large lubrication pressures, so let such a disk placed at a distance of about half its radius from the wall (such that $\epsilon= \delta -1 \sim0.5$) be torqued to rotate at $3\,\text{Hz}$ ($\Omega \sim 20\,\text{s}^{-1}$). The vertical velocity due to $\Pi$-dependence of viscosity is then $V\sim 5\,\upmu\text{m}/\text{s}$. This drift is manifestly irreversible: flipping the direction of rotation, for instance by imposing an oscillatory motion instead of constant rotation, does not reverse the direction of this drift and only continues to push the disk further away from the wall.

\section{Interfacial Magnus effect}

\begin{figure}
\centering
\includegraphics[width=\textwidth]{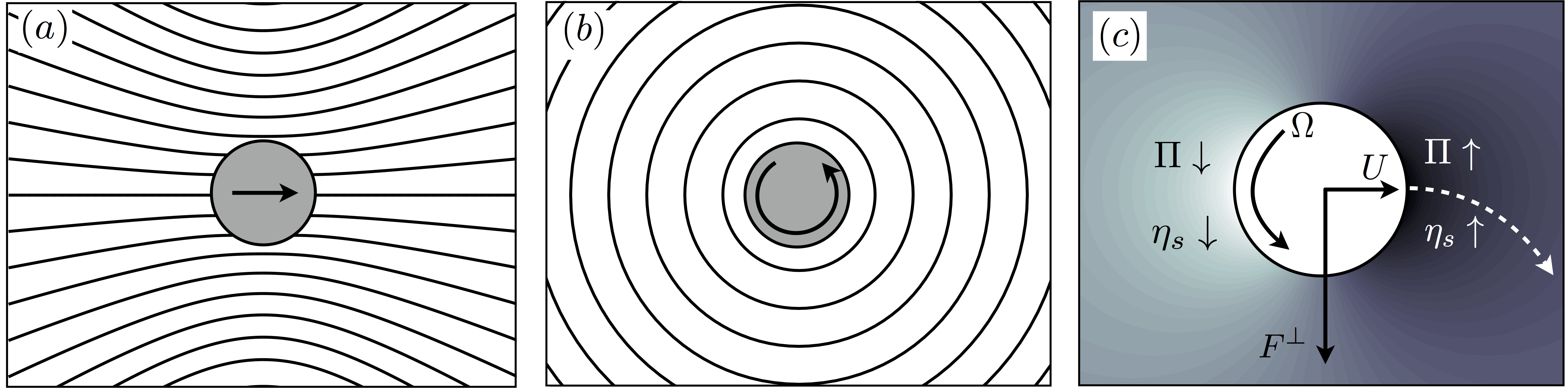}
\caption{Streamlines corresponding to a disk $(a)$ translating and $(b)$ rotating in an unbounded Newtonian monolayer. $(c)$ Surface pressure contours due to translation (rotation does not alter the surface pressure). Darker shades correspond to a relative increase in $\Pi$ ahead of the disk. The associated increase in $\eta_s$ leads to the $\oo{\zeta}$ force ${\bf F}^{\perp}$ that `rolls' the disk in a direction perpendicular to ${\bf U}$, leading to a curved trajectory. \label{fig:magnus}}
\end{figure}

More generally, the reciprocal theorem approach also applies to unconfined systems where Newtonian symmetries break as a result of $\Pi$-dependent viscosity. The integral in Eq.~\eqref{eq:correction} is non-linear in the $\oo{1}$ pressure and velocity fields, and coupled modes of motion can also lead to non-zero forces. We illustrate this using an interfacial analog of the Magnus effect in 3D fluids.

Consider the simultaneous translation and rotation of a disk of radius $a$ on an unbounded interface. The disturbance velocity (Fig.~\ref{fig:magnus}$(a)$) induced by its translation at velocity ${\bf U}$ in a Newtonian monolayer is a superposition of a 2D Stokeslet and a 2D potential dipole \cite{Pozrikidis}: 
\begin{equation}\label{eq:magnustrans}
\bu_t(\bx) = \left[ 2\left(-\ln(r){\bf I} +\frac{\bx\bx}{r^2} \right) +a^2\left(\frac{{\bf I}}{r^2}-\frac{2\bx\bx}{r^4}\right) \right] \bcdot {\bf U},
\end{equation}
where $\bx$ is measured from the center of the disk, and $r=|\bx|$. The associated surface pressure distribution is
\begin{equation}\label{eq:magnuspres}
\Pi_t(\bx)=\left[\frac{4\eta_s^0\bx}{r^2} \right]\bcdot{\bf U}.
\end{equation}
Similarly, the disturbance velocity corresponding to rotation (Fig.~\ref{fig:magnus}$(b)$) with angular velocity $\bOmega$ is
\begin{equation}\label{eq:magnusrot}
\bu_r(\bx) =\left(\frac{a^2}{r^2}\right) \bOmega \times \bx,
\end{equation}
while the surface pressure remains unchanged due to rotation alone: $\Pi_r(\bx)=0$.

These disturbance fields can be superimposed due to linearity of the $\oo{1}$ equations. As the auxiliary Newtonian problem in Eq.~\eqref{eq:correction}, we use the solution $\bu_{\rm aux}$ corresponding to a disk translating with velocity ${\bf U}_{\rm aux}$. The correction at $\oo{\zeta}$ then becomes
\begin{equation}\label{eq:magnusint}
{\bf F}_{1}\bcdot {\bf U}_{\rm aux}=- \int \Pi_t  \left[\del \left(\bu_t+\bu_r \right)+\left(\del \left(\bu_t +\bu_r \right)\right)^T\right]  \bcdotdot \del \bu_{\rm aux} \,dS.
\end{equation}
Neither translation nor rotation alone can generate a force at $\oo{\zeta}$ in an infinite $\Pi$-dependent monolayer, due to physical symmetries as described earlier. The integral in Eq.~\eqref{eq:magnusint} results in a non-zero force only due to the coupling between the two modes of motion. After some algebra, we find (in dimensional units)
\begin{equation}
{\bf F}^{\perp}\approx 16 \pi \frac{(\eta_s^0)^2}{\Pi_c}{\bf U} \times \bOmega,
\end{equation}
which acts perpendicular to the direction of motion.	

The importance of the coupling between translation and rotation is depicted in Fig.~\ref{fig:magnus}$(c)$. Translation sets up a surface pressure distribution such that $\Pi$ increases ahead of the disk and decreases in its rear. In a $\Pi$-thickening monolayer, $\eta_s(\Pi)$ changes in the same sense as $\Pi$. Consequently, the surface-viscous resistance to rotation is larger in front of the disk as compared to its rear. The disk responds by `rolling' on the more viscous side, moving in a direction perpendicular to ${\bf U}$, leading to a curved trajectory: Fig.~\ref{fig:magnus}$(c)$. The direction of the curve is opposite that of the traditional Magnus effect in 3D inertial fluids.

\section{Conclusion}

We have illustrated the leading-order symmetry-breaking consequences of surface-pressure-dependent surface viscosity, a feature of most insoluble surfactant monolayers. Working with incompressible monolayers with negligible bulk viscous stresses (`high-Boussinesq' limit), we applied the Lorentz reciprocal theorem to obtain perturbative corrections to Newtonian stresses when a disk is embedded in the monolayer. We focused primarily on bounded monolayers, which demonstrate the loss of Newtonian symmetries and reveal characteristic trends of the forces and torques generated by $\Pi$-dependent rheology. However, the formulation based on the reciprocal theorem is more general, as we illustrate with the interfacial Magnus effect.

We found that a disk translating parallel to (or rotating adjacent to) an interfacial barrier is subject to a force in the direction perpendicular to that barrier, leading to an evident loss of kinematic reversibility. The results are amplified in thin gaps, and are consistent with a lubrication theory adapted for $\eta_s(\Pi)$. As a striking display of kinematic irreversibility, we solved for the trajectory of a force-free disk rotated within a cylindrical cavity. In this 2D `journal bearing', forces perpendicular to the wall lead to unique spiral trajectories.

We have previously established the surprising consequences of $\Pi$-dependent viscosity in small fluid gaps \cite{Manikantan2017}, for any form of  $\eta_s(\Pi)$ and arbitrary velocities. The current work, on the other hand, focuses on `small' velocities in a perturbative sense, without restrictions on the thickness of the fluid gap. Between these two works, we now have a clearer grasp of the impact of $\Pi$-dependent rheology, with quantitative predictions and qualitative insights into practically relevant situations. With the rising prominence and success of interfacial microrheology \cite{Choi2011,Zell2014,Verwijlen2011,Hermans2014}, these results are of particular experimental relevance due to their implications on the measurement and interpretation of surface rheology. 

Finally, coupling different modes of motion can potentially lead to non-intuitive trajectories, as evinced by the example of the interfacial Magnus effect. This also applies to the translation of a disk in a background surface flow with a rotational component, or more broadly to the interaction between two particles embedded in a $\Pi$-thickening or -thinning monolayer. Such a pair problem, like its 3D analogue \cite{Batchelor1972}, sets the stage for a hydrodynamic theory of the rheology of 2D suspensions. Changes in local microstructure, guided by irreversible pair interactions due to $\eta_s(\Pi)$, could lead to hydrodynamic aggregation or separation of particles, which in turn result in macroscopic shear-thickening or -thinning. The mathematical machinery developed here, and the qualitative insights gained, will inform these studies, which we leave for future work.

\section{Acknowledgments}
This work was supported by the National Science Foundation (NSF) under Grant No. CBET-1512833. The authors thank Ian Williams for a critical reading of the manuscript.

\appendix
\section{Stream function coefficients}\label{sec:appendix}
Here we list the coefficients in the series solution to the stream function in Eq.~\eqref{eq:bihar_soln} for the general case of a cylinder within a cylinder enclosing a 2D Newtonian fluid. The translating and/or rotating inner cylinder will be identified by the streamline corresponding to $\alpha_0$, while the stationary outer wall is identified by $\alpha_1$ (see Fig.~\ref{fig:geometry}$(b)$). In every case, $\alpha_1=0$ retrieves the limit of a plane wall along $y=0$. We provide the coefficients corresponding to each mode of motion separately: since these are solutions to the Stokes equations, they may be suitably superimposed.

\subsection{Translation perpendicular to the wall}
For radial motion of the inner cylinder towards the center of the outer cylinder with velocity $V$ (or equivalently, motion along the positive $y$-axis in Fig.~\ref{fig:geometry}$(b)$), the only non-zero coefficients are
\begin{subequations}
\begin{align}
D&=\frac{V}{\mathcal{M}}\\
a_1'&=-\frac{1}{2} \frac{V}{\mathcal{M}} \frac{\sinh (\alpha_0+\alpha_1)}{\cosh(\alpha_0-\alpha_1)},\\
b_1'&=\frac{1}{2} \frac{V}{\mathcal{M}} \frac{\cosh (\alpha_0+\alpha_1) }{\cosh(\alpha_0-\alpha_1)},\\
c_1'&=\frac{1}{2} \frac{V}{\mathcal{M}} (\tanh (\alpha_0-\alpha_1)+2 \alpha_1),
\end{align}
\end{subequations}
where
\begin{equation}
\mathcal{M}=\tanh (\alpha_0-\alpha_1)-(\alpha_0-\alpha_1).
\end{equation}

\subsection{Translation parallel to wall}
For azimuthal translation of the inner cylinder velocity $U$ (or equivalently, motion along the positive $x$-axis in Fig.~\ref{fig:geometry}$(b)$), the only non-zero coefficients are
\begin{subequations}
\begin{align}
A&=\frac{U}{\mathcal{L}} \left[\sinh \left(2 \alpha _0\right)+\sinh \left(2 \alpha _1\right)\right],\\
B&=\frac{U}{\mathcal{L}} \left[\alpha _1\left\{\cosh \left(2 \alpha _0\right)+\cosh \left(2 \alpha _1\right)-2\right\}- \cosh \left(\alpha _1\right) \left\{\sinh \left(2 \alpha _0-\alpha _1\right)+3 \sinh \left(\alpha _1\right)\right\}\right],\\
C&=-\frac{U}{\mathcal{L}} \left[\cosh \left(2 \alpha _0\right)+\cosh \left(2 \alpha _1\right)-2\right],\\
a_1&=-\frac{U}{\mathcal{L}} \sinh ^2\left(\alpha _0+\alpha _1\right),\\
b_1&=\frac{1}{2} \frac{U}{\mathcal{L}} \sinh \left(2 \left(\alpha _0+\alpha _1\right)\right),\\
c_1&=\frac{1}{2} \frac{U}{\mathcal{L}} \left[2 \cosh \left(2 \alpha _0\right)+2 \cosh \left(2 \alpha _1\right)-\cosh \left(2 \left(\alpha _0-\alpha _1\right)\right)-3\right],
\end{align}
\end{subequations}
with
\begin{equation}
\mathcal{L}=(\alpha _0-\alpha_1) \left[\cosh \left(2 \alpha _0\right)+\cosh \left(2 \alpha _1\right)-2\right]-4 \sinh \left(\alpha _0\right) \sinh \left(\alpha _1\right) \sinh \left(\alpha _0-\alpha _1\right) .
\end{equation}

\subsection{Rotation next to wall}
For rotation of inner cylinder towards the center of the outer cylinder with angular velocity $\Omega$ (such that $\Omega$ is positive in the counter-clockwise sense in Fig.~\ref{fig:geometry}$(b)$), the only non-zero coefficients are
\begin{subequations}
\begin{align}
A&=-\frac{a\Omega}{\mathcal{N}} \coth \left(\alpha _0-\alpha _1\right) \left[2 \left(\alpha _0-\alpha _1\right) \sinh \left(\alpha _0\right)-\cosh \left(\alpha _0\right)+\cosh \left(\alpha _0-2 \alpha _1\right)\right],\\
B&=-\frac{a\Omega}{\mathcal{N}} \left[\cosh \left(\alpha _0\right)\left\{2 \alpha _1+\sinh \left(2 \alpha _1\right)\right\} \right. \nonumber \\
& \quad \quad \left. +\coth \left(\alpha _0-\alpha _1\right) \left\{\alpha _1 \sinh \left(2 \alpha _1\right) \cosh \left(\alpha _0\right)-2 \alpha _0 \sinh \left(\alpha _0\right) \cosh ^2\left(\alpha _1\right)\right\}\right],\\
C&=2 \frac{a\Omega}{\mathcal{N}} \sinh \left(\alpha _0-\alpha _1\right) \sinh \left(\alpha _1\right),\\
a_1&=-\frac{a\Omega}{\mathcal{N}} \sinh \left(\alpha _0+\alpha _1\right) \left[\sinh \left(\alpha _1\right)+\left(\alpha _1-\alpha _0\right) \sinh \left(\alpha _0\right) \text{csch}\left(\alpha _0-\alpha _1\right)\right],\\
b_1&=-\frac{1}{2} \frac{a\Omega}{\mathcal{N}} \frac{\cosh \left(\alpha _0+\alpha _1\right)}{\sinh\left(\alpha _0-\alpha _1\right)} \left[2 \left(\alpha _0-\alpha _1\right) \sinh \left(\alpha _0\right)-\cosh \left(\alpha _0\right)+\cosh \left(\alpha _0-2 \alpha _1\right)\right],\\
c_1&=-\frac{1}{2} \frac{a\Omega}{\mathcal{N}} \left[2 \left(\alpha _0-\alpha _1\right) \sinh \left(\alpha _0\right) \left\{\sinh \left(2 \alpha _1\right) \coth \left(\alpha _0-\alpha _1\right)+1\right\} \right. \nonumber \\
& \hspace{2.5in} \left. -2 \sinh \left(\alpha _1\right) \sinh \left(\alpha _0+\alpha _1\right)\right],
\end{align}
\end{subequations}
where $\mathcal{L}$ is the same as with parallel translation.


\begin{thebibliography}{31}
\expandafter\ifx\csname urlstyle\endcsname\relax
  \providecommand{\doi}[1]{doi:\discretionary{}{}{}#1}\else
  \providecommand{\doi}{doi:\discretionary{}{}{}\begingroup
  \urlstyle{rm}\Url}\fi

\bibitem{ScrivenSternling}
Scriven, L.~E. \& Sterling, C.~V., 1960 {The Marangoni Effects}.
\newblock \emph{Nature} \textbf{187}, 186--188.
\newblock (\doi{10.1038/187186a0}).

\bibitem{Leal}
Leal, L.~G., 2007 \emph{{Advanced Transport Phenomena}}.
\newblock Cambridge University Press.

\bibitem{Fuller2012}
Fuller, G.~G. \& Vermant, J., 2012 {Complex Fluid-Fluid Interfaces: Rheology
  and Structure}.
\newblock \emph{Annual Review of Chemical and Biomolecular Engineering}
  \textbf{3}, 519--543.
\newblock (\doi{10.1146/annurev-chembioeng-061010-114202}).

\bibitem{Langevin2014}
Langevin, D., 2014 {Rheology of Adsorbed Surfactant Monolayers at Fluid
  Surfaces}.
\newblock \emph{Annual Review of Fluid Mechanics} \textbf{46}, 47--65.
\newblock (\doi{10.1146/annurev-fluid-010313-141403}).

\bibitem{Levich}
Levich, V.~G., 1962 \emph{{Physicochemical Hydrodynamics}}.
\newblock Prentice Hall.

\bibitem{Fischer2004}
Fischer, T.~M., 2004 {Comment on ``Shear Viscosity of Langmuir Monolayers in
  the Low-Density Limit''}.
\newblock \emph{Physical Review Letters} \textbf{92}, 139603.
\newblock (\doi{10.1103/PhysRevLett.92.139603}).

\bibitem{Choi2011}
Choi, S.~Q., Steltenkamp, S., Zasadzinski, J.~A. \& Squires, T.~M., 2011
  {Active microrheology and simultaneous visualization of sheared phospholipid
  monolayers.}
\newblock \emph{Nature communications} \textbf{2}, 312.
\newblock (\doi{10.1038/ncomms1321}).

\bibitem{Zell2014}
Zell, Z.~A., Nowbahar, A., Mansard, V., Leal, L.~G., Deshmukh, S.~S., Mecca,
  J.~M., Tucker, C.~J. \& Squires, T.~M., 2014 {Surface shear inviscidity of
  soluble surfactants}.
\newblock \emph{Proceedings of the National Academy of Sciences} \textbf{111},
  3677--3682.
\newblock (\doi{10.1073/pnas.1315991111}).

\bibitem{Brooks1999}
Brooks, C.~F., Fuller, G.~G., Frank, C.~W. \& Robertson, C.~R., 1999 An
  interfacial stress rheometer to study rheological transitions in monolayers
  at the air-water interface.
\newblock \emph{Langmuir} \textbf{15}, 2450--2459.
\newblock (\doi{10.1021/la980465r}).

\bibitem{Verwijlen2011}
Verwijlen, T., Moldenaers, P., Stone, H.~A. \& Vermant, J., 2011 {Study of the
  Flow Field in the Magnetic Rod Interfacial Stress Rheometer}.
\newblock \emph{Langmuir} \textbf{27}, 9345--9358.
\newblock (\doi{10.1021/la201109u}).

\bibitem{Saffman1975}
Saffman, P.~G. \& Delbruck, M., 1975 {Brownian motion in biological membranes.}
\newblock \emph{Proceedings of the National Academy of Sciences} \textbf{72},
  3111--3113.
\newblock (\doi{10.1073/pnas.72.8.3111}).

\bibitem{Saffman1976}
Saffman, P.~G., 1976 {Brownian motion in thin sheets of viscous fluid}.
\newblock \emph{Journal of Fluid Mechanics} \textbf{73}, 593.
\newblock (\doi{10.1017/S0022112076001511}).

\bibitem{Prasad2006}
Prasad, V., Koehler, S.~A. \& Weeks, E.~R., 2006 {Two-Particle Microrheology of
  Quasi-2D Viscous Systems}.
\newblock \emph{Physical Review Letters} \textbf{97}, 176001.
\newblock (\doi{10.1103/PhysRevLett.97.176001}).

\bibitem{Scriven1960}
Scriven, L., 1960 {Dynamics of a fluid interface Equation of motion for
  Newtonian surface fluids}.
\newblock \emph{Chemical Engineering Science} \textbf{12}, 98--108.
\newblock (\doi{10.1016/0009-2509(60)87003-0}).

\bibitem{Bair1998}
Bair, S., Khonsari, M. \& Winer, W.~O., 1998 {High-pressure rheology of
  lubricants and limitations of the Reynolds equation}.
\newblock \emph{Tribology International} \textbf{31}, 573--586.
\newblock (\doi{10.1016/S0301-679X(98)00078-4}).

\bibitem{Hron2001}
Hron, J., Malek, J. \& Rajagopal, K.~R., 2001 {Simple flows of fluids with
  pressure-dependent viscosities}.
\newblock \emph{Proceedings of the Royal Society A: Mathematical, Physical and
  Engineering Sciences} \textbf{457}, 1603--1622.
\newblock (\doi{10.1098/rspa.2000.0723}).

\bibitem{Rajagopal2003}
Rajagopal, K.~R. \& Szeri, A.~Z., 2003 {On an inconsistency in the derivation
  of the equations of elastohydrodynamic lubrication}.
\newblock \emph{Proceedings of the Royal Society A: Mathematical, Physical and
  Engineering Sciences} \textbf{459}, 2771--2786.
\newblock (\doi{10.1098/rspa.2003.1145}).

\bibitem{Denn1981}
Denn, M.~M., 1981 {Pressure drop-flow rate equation for adiabatic capillary
  flow with a pressure- and temperature-dependent viscosity}.
\newblock \emph{Polymer Engineering and Science} \textbf{21}, 65--68.
\newblock (\doi{10.1002/pen.760210202}).

\bibitem{Penwell1971}
Penwell, R.~C., Porter, R.~S. \& Middleman, S., 1971 {Determination of the
  pressure coefficient and pressure effects in capillary flow}.
\newblock \emph{Journal of Polymer Science Part A-2: Polymer Physics}
  \textbf{9}, 731--745.
\newblock (\doi{10.1002/pol.1971.160090412}).

\bibitem{Kaganer1999}
Kaganer, V., M{\"{o}}hwald, H. \& Dutta, P., 1999 {Structure and phase
  transitions in Langmuir monolayers}.
\newblock \emph{Reviews of Modern Physics} \textbf{71}, 779--819.
\newblock (\doi{10.1103/RevModPhys.71.779}).

\bibitem{Kurtz2006}
Kurtz, R.~E., Lange, A. \& Fuller, G.~G., 2006 Interfacial rheology and
  structure of straight-chain and branched fatty alcohol mixtures.
\newblock \emph{Langmuir} \textbf{22}, 5321--5327.
\newblock (\doi{10.1021/la060290i}).

\bibitem{Kim2011}
Kim, K., Choi, S.~Q., Zasadzinski, J.~A. \& Squires, T.~M., 2011 {Interfacial
  microrheology of DPPC monolayers at the air-water interface}.
\newblock \emph{Soft Matter} \textbf{7}, 7782.
\newblock (\doi{10.1039/c1sm05383c}).

\bibitem{Kim2013}
Kim, K., Choi, S.~Q., Zell, Z.~A., Squires, T.~M. \& Zasadzinski, J.~A., 2013
  {Effect of cholesterol nanodomains on monolayer morphology and dynamics}.
\newblock \emph{Proceedings of the National Academy of Sciences} \textbf{110},
  E3054--E3060.
\newblock (\doi{10.1073/pnas.1303304110}).

\bibitem{Hermans2014}
Hermans, E. \& Vermant, J., 2014 {Interfacial shear rheology of DPPC under
  physiologically relevant conditions}.
\newblock \emph{Soft Matter} \textbf{10}, 175--186.
\newblock (\doi{10.1039/C3SM52091A}).

\bibitem{Manikantan2017}
Manikantan, H. \& Squires, T.~M., 2017 {Pressure-dependent surface viscosity
  and its surprising consequences in interfacial lubrication flows}.
\newblock \emph{Physical Review Fluids} \textbf{2}, 023301.
\newblock (\doi{10.1103/PhysRevFluids.2.023301}).

\bibitem{KimKarrila}
Kim, S. \& Karrila S.~J., 1991 \emph{{Microhydrodynamics: principles and selected applications}}.
\newblock Butterworth-Heinemann.

\bibitem{Wakiya1975}
Wakiya, S., 1975 {Application of Bipolar Coordinates to the Two-Dimensional
  Creeping Motion of a Liquid. II. Some Problems for Two Circular Cylinders in
  Viscous Fluid}.
\newblock \emph{Journal of the Physical Society of Japan} \textbf{39},
  1603--1607.
\newblock (\doi{10.1143/JPSJ.39.1603}).

\bibitem{Jeffrey1981}
Jeffrey, D.~J. \& Onishi, Y., 1981 {The slow motion of a cylinder next to a
  plane wall}.
\newblock \emph{Quarterly Journal of Mechanics and Applied Mathematics}
  \textbf{34}, 129--137.
\newblock (\doi{10.1093/qjmam/34.2.129}).

\bibitem{Keh1985}
Keh, H.~J. \& Anderson, J.~L., 1985 {Boundary effects on electrophoretic motion
  of colloidal cylinders}.
\newblock \emph{Journal of Fluid Mechanics} \textbf{153}, 417.
\newblock (\doi{10.1017/S002211208500132X}).

\bibitem{Darabaner1967}
Darabaner, C.~L., Raasch, J.~K. \& Mason, S.~G., 1967 {Particle motions in
  sheared suspensions XX: Circular cylinders}.
\newblock \emph{The Canadian Journal of Chemical Engineering} \textbf{45},
  3--12.
\newblock (\doi{10.1002/cjce.5450450102}).

\bibitem{Pozrikidis}
Pozrikidis, C., 1997 \emph{{Introduction to Theoretical and Computational Fluid
  Dynamics}}.
\newblock Oxford University Press.

\bibitem{Batchelor1972}
Batchelor, G.~K. \& Green, J.~T., 1972 {The hydrodynamic interaction of two
  small freely-moving spheres in a linear flow field}.
\newblock \emph{Journal of Fluid Mechanics} \textbf{56}, 375.
\newblock (\doi{10.1017/S0022112072002927}).

\end{thebibliography}

\end{document}